\documentclass[epj]{svjour2}

\usepackage{epsfig}

\def\p{I\!\!P}

\def\beq{\begin{equation}}
\def\eeq{\end{equation}}
\def\bea{\begin{eqnarray}}
\def\eea{\end{eqnarray}}

\begin{document}

\title{Factorization Breaking in Diffractive Photoproduction of Dijets}

\author{M.\ Klasen\inst{1,2} \and
 G.\ Kramer\inst{2}}

\institute{Laboratoire de Physique Subatomique et de Cosmologie,
 Universit\'e Joseph Fourier/CNRS-IN2P3, 53 avenue
 des Martyrs, F-38026 Grenoble, France, \email{klasen@lpsc.in2p3.fr}
 \and
 II.\ Institut f\"ur Theoretische Physik, Universit\"at Hamburg,
 Luruper Chaussee 149, D-22761 Hamburg, Germany}

\date{Date: \today}

\abstract{
We have calculated the diffractive dijet cross section in low-$Q^2$ $ep$
scattering in the HERA regime. The results of the calculation in LO and
NLO are compared to recent experimental data of the H1 collaboration.
We find that in LO the calculated cross sections are in reasonable agreement
with the experimental results. In NLO, however, some of the cross sections
disagree, showing that factorization breaking occurs in that order. By
suppressing the resolved contribution by a factor of approximately three,
good agreement with all the data is found. The size of the factorization
breaking effects in diffractive dijet photoproduction agrees well with
absorptive model predictions.
}

\maketitle


\section{Introduction}
\label{sec:1}

Diffractive $\gamma p$ interactions are characterized by an outgoing proton
of high longitudinal momentum and/or a large rapidity gap, defined as a
region of pseudo-rapidity, $\eta = -\ln \tan \theta/2 $, devoid of
particles. It is assumed that the large rapidity gap is due to the exchange
of a pomeron, which carries the internal quantum numbers of the vacuum.
Diffractive events that contain a hard scattering are referred to as hard
diffraction. A necessary condition for a hard scattering is the occurrence
of a hard scale, which may be the large momentum transfer $Q^2$ in inclusive
deep-inelastic $ep$ scattering, the high transverse momentum of jets or
single hadrons, or the mass of heavy quarks or of $W$-bosons produced in
high-energy $\gamma p$, $ ep$ or $p\bar{p}$ collisions.

The central problem in hard diffraction is the question of QCD
factorization, {\it i.e.} the question whether it is possible to explain the
observed cross sections in hard diffractive processes by a convolution of
diffractive parton distribution functions (PDFs) with parton-level cross
sections.

The diffractive PDFs have been determined by the H1 collaboration from a
recent high-precision inclusive measurement of the diffractive deep
inelastic scattering (DIS) process $ep \rightarrow eXY$, where $Y$ is a
single proton or a low mass proton excitation \cite{H1}. The diffractive
PDFs can serve as input for the calculation of any of the other diffractive
hard scattering reactions mentioned above. For diffractive DIS, QCD
factorization has been proven by Collins \cite{Coll}. This has the
consequence that the evolution of the diffractive PDFs is predictable in the
same way as the PDFs of the proton via the DGLAP evolution equations.
Collins' proof is valid for all lepton-induced collisions. These include
besides diffractive DIS also the diffractive direct photoproduction of jets.
The proof fails for hadron-induced processes.

As is well known, the cross section for the photoproduction of jets
is the sum of the direct contribution, where the photon couples directly
to the quarks, and of the resolved contribution, where the photon first
resolves into partons (quarks or gluons), which subsequently induce the
hard scattering to produce the jets in the final state. So, the resolved
part resembles hadron-induced production of jets as for example in
$p\bar{p}$ collisions. Dijet production in single-diffractive collisions has
been measured recently by the CDF collaboration at the Tevatron \cite{CDF}.
It was found that the dijet cross section was suppressed relative to the
prediction based on older diffractive PDFs from the H1 collaboration
\cite{H11} by one order of magnitude \cite{CDF}. From this result we would
conclude that the resolved contribution in diffractive photoproduction of
jets should be reduced by a correction factor similar to the one needed in
hadron-hadron scattering \cite{Al}. This suppression factor (sometimes also
called the rapidity gap survival probability) has been calculated using
various eikonal models, based on multi-pomeron exchanges and $s$-channel
unitarity \cite{Kaidalov:2001iz}. The direct and the resolved parts of
the cross section contribute with varying strength in different kinematic
regions. In particular, the $x_{\gamma}$-distribution is very sensitively 
dependent on the way how these two parts of the cross section are 
superimposed. Near $x_{\gamma } \simeq 1$ the direct part dominates,
whereas for $x_{\gamma } < 1$ the resolved part gives the main
contribution. However, in this region also contributions from
next-to-leading order (NLO) corrections of the direct cross section occur.
Therefore, to decide whether the resolved part is suppressed as compared to
the experimental data, a NLO analysis is actually needed. This is the aim of
this paper. For our calculations we rely on our work on dijet production in
the inclusive (sum of diffractive and non-diffractive) reaction $\gamma + p
\rightarrow {\rm jets}+X$ \cite{KKK}, in which we have calculated the cross 
sections for inclusive one-jet and two-jet production up to NLO for both
the direct and the resolved contribution. The predictions of this and other
work \cite{Frixione:1997ks} have been tested now by many experimental
studies of the H1 and ZEUS collaborations \cite{H12,ZEUS}. Very good
agreement with the experimental data \cite{H12,ZEUS} has been found. From
these comparisons it follows that a leading order (LO) calculation is not
sufficient. It underestimates the measured cross section by up to $50\%$
\cite{Klasen:2002xb}.

The question whether the resolved cross section needs a suppression
factor, can be decided first by looking at the shape of those
distributions which are particularly sensitively dependent on the resolved
contributions, as for example the $x_\gamma$-distribution for the smaller
$x_\gamma$ or the $E_T$-distributions at small $E_T$. Because of the 
interplay of direct and resolved contributions, LO calculations are not 
sufficient, in particular, since the NLO corrections are much more important
for the resolved than the direct part. This is even more
important if one looks at the normalization of the differential cross
sections.

Recently the H1 collaboration \cite{H13} have presented data for
differential dijet cross sections in the low-$|t|$ diffractive
photoproduction process $ep \rightarrow eXY$, in which the photon
dissociation system $X$ is separated from a leading low-mass baryonic system
$Y$ by a large rapidity gap. Using the same kinematic constraint as in these
measurements we shall calculate the same cross section as in the H1 analysis
up to NLO. By comparing to the data we shall try to find out, whether or not
a suppression of the resolved cross section is needed in order to find
reasonable agreement between the data and the theoretical predictions.

The outline of this work is as follows. In Sec.\ \ref{sec:2}, we specify the
kinematic variables used in the analysis and describe the input for the
calculation of the diffractive dijet cross section. In Sec.\ \ref{sec:3}, we
report our results and discuss our findings concerning the suppression
factor for the resolved contributions. Section \ref{sec:4} contains our
conclusions and the outlook to further work.

\section{Kinematic Variables and Diffractive Parton Distributions}
\label{sec:2}

\subsection{Kinematic Variables and Constraints}

\begin{figure}
 \centering
 \epsfig{file=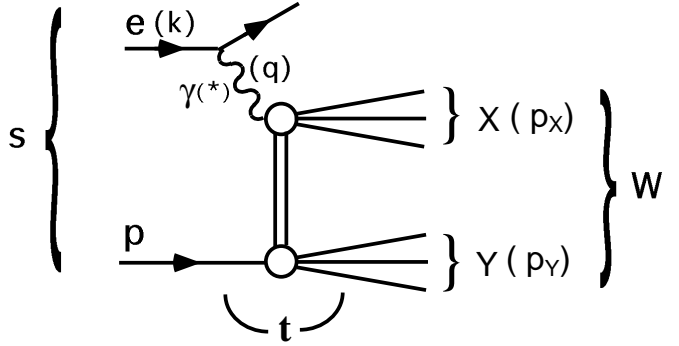,width=\columnwidth}
 \caption{\label{fig:1}Diffractive scattering process $ep\to eXY$, where
 the hadronic systems $X$ and $Y$ are separated by the largest rapidity
 gap in the final state.}
\end{figure}

The diffractive process $ep \rightarrow eXY$, in which the systems $X$ and 
$Y$ are separated by the largest rapidity gap in the final state, is
sketched in Fig.~\ref{fig:1}.
%
%
The system $X$ contains at least two jets, and the system $Y$ is 
supposed to be a proton or another low-mass baryonic system. Let $k$ and $p$
denote the momenta of the incoming electron (or positron) and proton, 
respectively and $q$ the momentum of the virtual photon $\gamma ^{*}$. Then 
the usual kinematic variables are
\begin{equation}
 s = (k+p)^2,~ Q^2 = -q^2,~{\rm and}~y = \frac{qp}{kp}.
\end{equation}
We denote the four-momenta of the systems $X$ and $Y$ by $p_X$ and $p_Y$. 
The H1 data \cite{H13} are described in terms of
\begin{eqnarray}
 M_X^2 = p_X^2 & ~{\rm and}~ & t = (p-p_Y)^2, \nonumber\\
 M_Y^2 = p_Y^2 & ~{\rm and}~ & x_{\p} = \frac{q(p-p_Y)}{qp},
\end{eqnarray}
where $M_X$ and $M_Y$ are the invariant masses of the systems $X$ and $Y$,
$t$ is the squared four-momentum transfer of the incoming proton and the
system $Y$, and $x_{\p}$ is the momentum fraction of the proton beam
transferred to the system $X$.

The exchange between the systems $X$ and $Y$ is supposed to be the pomeron 
$\p$ or any other Regge pole, which couples to the proton and the system
$Y$ with four-momentum $p-p_Y$. The pomeron is resolved into partons (quarks
or gluons) with four-momentum $v$. In the same way the virtual photon can
resolve into partons with four-momentum $u$, which is equal to $q$ for the
direct process. With these two momenta $u$ and $v$ we define
\begin{equation}
 x_{\gamma } = \frac{pu}{pq} ~{\rm and}~ z_{\p} = \frac{qv}{q(p-p_Y)}.
\end{equation}
$x_\gamma$ is the longitudinal momentum fraction carried by the partons 
coming from the photon, and $z_{\p}$ is the corresponding quantity carried
by the partons of the pomeron etc., {\it i.e.} the diffractive exchange. For
the direct process we have $x_{\gamma } = 1$. The final state, produced by
the ingoing momenta $u$ and $v$, has the invariant mass $M_{12} =
\sqrt{(u+v)^2}$, which is equal to the invariant dijet mass in the case that
no more than two hard jets are produced. $q-u$ and $p-p_Y-v$ are the
four-momenta of the remnant jets produced at the photon and pomeron side.
The regions of the kinematic variables, in which the cross section has been
measured by the H1 collaboration \cite{H13}, are given in Tab.~\ref{tab:1}.
\begin{table}
 \caption{\label{tab:1}Regions of kinematic variables.}
 \begin{center}
 \begin{tabular}{rcccl}
 \hline\noalign{\smallskip}
  0.3   & $<$ & $y$   & $<$ & 0.65 \\
        &     & $Q^2$ & $<$ & 0.01 GeV$^2$ \\
        &     & $E_T^{\rm jet1}$ & $>$ & 5 GeV \\
        &     & $E_T^{\rm jet2}$ & $>$ & 4 GeV \\
  $-1$  & $<$ & $\eta_{\rm lab}^{\rm jet1,2}$ & $<$ & 2 \\
        &     & $x_{\p}$ & $<$ & 0.03 \\
        &     & $M_Y$ & $<$ & 1.6 GeV \\
        &     & $-t$  & $<$ & 1 GeV$^2$ \\
 \noalign{\smallskip}\hline
 \end{tabular}
 \end{center}
\end{table}
With the same constraints we have evaluated the theoretical cross sections.

The upper limit of $x_{\p}$ is kept small in order for the pomeron exchange
to be dominant. In the experimental analysis as well as in the NLO
calculations, jets are defined with the inclusive $k_T$-cluster algorithm
with a distance parameter $d=1$ \cite{ES} in the laboratory frame. At least
two jets are required with transverse energies $E_T^{\rm jet1} > 5$ GeV and 
$E_T^{\rm jet2} > 4$ GeV. They are the leading and subleading jets with 
$-1<\eta_{\rm lab}^{\rm jet1,2} < 2$. The lower limits of the jet $E_T$'s
are asymmetric in order to avoid infrared sensitivity in the computation of
the NLO cross sections, which are integrated over $E_T$ \cite{KK}.

In the experimental analysis the variable $y$ is deduced from the energy 
$E_e'$ of the scattered electron $y=1-E_e'/E_e$. Furthermore,
$sy=W^2=(q+p)^2=(p_X+p_Y)^2$. $x_{\p}$ is reconstructed according to
\begin{equation}
 x_{\p} = \frac{\sum_{X} (E+p_z)}{2E_p},
\end{equation}
where $E_p$ is the proton beam energy and the sum runs over all particles
(jets) in the $X$-system. The variables $M_{12}$, $x_{\gamma }$, and
$z_{\p}$ are determined only from the kinematic variables of the two hard
leading jets with four-momenta $p^{\rm jet1}$ and $p^{\rm jet2}$. So,
\begin{equation}
 M_{12}^2 = (p^{\rm jet1} + p^{\rm jet2})^2
\end{equation}
where additional jets are not taken into account. In the same way
\begin{equation}
 x_{\gamma }^{\rm jets} =\frac{\sum_{\rm jets}(E-p_z)}{2yE_e} ~{\rm and}~
 z_{\p}^{\rm jets} = \frac{\sum_{\rm jets}(E+p_z)}{2x_{\p} E_p}.
\end{equation}
The sum over jets runs only over the variables of the two leading jets.
These definitions for $x_{\gamma }$ and $z_{\p}$ are not the same as the
definitions given earlier, where also the remnant jets and any additional
hard jets are taken into account in the final state. In the same way $M_X$
can be estimated by $M_X^2 = M_{12}^2/(z_{\p}^{\rm jets} x_{\gamma}^{\rm
jets})$. The dijet system is characterized by the transverse energies
$E_T^{\rm jet1}$ and $E_T^{\rm jet2}$ and the rapidities in the laboratory
system $\eta^{\rm jet1}_{\rm lab}$ and $\eta^{\rm jet2}_{\rm lab}$. The
differential cross sections are measured and calculated as functions of the
transverse energy $E_T^{\rm jet1}$ of the leading jet, the average rapidity
$\overline{\eta}^{\rm jets} = (\eta^{\rm jet1}_{\rm lab} + \eta^{\rm
jet2}_{\rm lab})/2$, and the jet separation $|\Delta\eta^{\rm jets}| =
|\eta^{\rm jet1}_{\rm lab} - \eta^{\rm jet2}_{\rm lab}|$, which is related
to the scattering angle in the center-of-mass system of the two hard jets.

\subsection{Diffractive Parton Distributions}

The diffractive PDFs are obtained from an analysis of the diffractive
process $ep \rightarrow eXY$, which is illustrated in Fig.~\ref{fig:1},
where now $Q^2$ is large and the state $X$ consists of all possible final
states, which are summed. The cross section for this diffractive DIS process
depends in general on five independent variables (azimuthal angle dependence
neglected): $Q^2$, $x$ (or $\beta$), $x_{\p}$, $M_Y$, and $t$. These
variables are defined as before, and $x= Q^2/(2pq) = Q^2/(Q^2+W^2) = x_{\p}
\beta$. The system $Y$ is not measured, and the results are integrated over
$-t < 1$ GeV$^2$ and $M_Y < 1.6$ GeV as in the photoproduction case. The
measured cross section is expressed in terms of a reduced diffractive cross
section $\sigma^{D(3)}_r$ defined through
\begin{equation}
 \frac{d^3 \sigma^D}{dx_{\p}dxdQ^2} = \frac{4\pi \alpha^2}{xQ^4}
 \left(1-y+\frac{y^2}{2}\right) \sigma^{D(3)}_r(x_{\p},x,Q^2)
\end{equation}
and is related to the diffractive structure functions $F^{D(3)}_2$ and
$F^{D(3)}_L$ by
\begin{equation}
 \sigma^{D(3)}_r = F^{D(3)}_2 - \frac{y^2}{1+(1-y)^2} F^{D(3)}_L.
\end{equation}
$y$ is defined as before, and $F^{D(3)}_L$ is the longitudinal diffractive
structure function.

The proof of Collins \cite{Coll}, that QCD factorization is applicable to
diffractive DIS, has the consequence that the DIS cross section for 
$\gamma ^{*}p \rightarrow XY$ can be written as a convolution of a partonic 
cross section $\sigma^{\gamma ^{*}}_a$, which is calculable as an expansion 
in the strong coupling constant $\alpha_s$, with diffractive PDFs $f^D_a$ 
yielding the probability distribution for a parton $a$ in the proton under
the constraint that the proton undergoes a scattering with a particular
value for the squared momentum transfer $t$ and $x_{\p}$. Then the cross
section for $\gamma^{*} p \rightarrow XY$ is
\begin{equation} 
 \frac{d^2\sigma }{dx_{\p}dt} = \sum_{a} \int_{x}^{x_{\p}} d\xi
 \sigma^{\gamma *}_{a}(x,Q^2,\xi) f_a^D(\xi,Q^2;x_{\p},t).
\end{equation}
This formula is valid for sufficiently large $Q^2$ and fixed $x_{\p}$ and
$t$. The parton cross sections are the same as those for inclusive DIS. The
diffractive PDFs are non-perturba\-tive objects. Only their $Q^2$ evolution
can be predicted with the well known DGLAP evolution equations, which we
shall use in LO and NLO.

Usually for $f_a^D(x,Q^2;x_{\p},t)$ an additional assumption is made, namely
that it can be written as a product of two factors, $f_{\p/p}(x_{\p},t)$ and
$f_{a/\p}(\beta,Q^2)$,
\begin{equation}
 f_a^D(x,Q^2;x_{\p},t) = f_{\p/p}(x_{\p},t) f_{a/\p}(\beta=x/x_{\p},Q^2).
\end{equation}

$f_{\p/p}(x_{\p},t)$ is the pomeron flux factor. It gives the probability
that a pomeron with variables $x_{\p}$ and $t$ couples to the proton. Its
shape is controlled by Regge asymptotics and is in principle measurable by
soft processes under the condition that they can be fully described by
single pomeron exchange. This Regge factorization formula, first introduced
by Ingelman and Schlein \cite{JS}, represents the resolved pomeron model, in
which the diffractive exchange, {\it i.e.} the pomeron, can be considered as
a quasi-real particle with a partonic structure given by PDFs $f_{a/\p}
(\beta,Q^2)$. $\beta$ is the longitudinal momentum fraction of the pomeron
carried by the emitted parton $a$ in the pomeron. The important point is
that the dependence of $f_a^D$ on the four variables $x, Q^2, x_{\p}$ and
$t$ factorizes in two functions $f_{\p/p}$ and $f_{a/\p}$, which each depend
only on two variables.

Since the value of $t$ could not be fixed in the diffractive DIS
measurements, it has been integrated over with $t$ varying in the region
$t_{\rm cut} < t < t_{\min}$. Therefore we have according to \cite{H1}
\begin{equation}
 f(x_{\p}) = \int_{t_{\rm cut}}^{t_{\min}} dt f_{\p/p}(x_{\p},t),
\end{equation}
where $t_{\rm cut} = -1$ GeV$^2$ and $t_{\min}$ is the minimum kinematically
allowed value of $|t|$. In \cite{H1} the pomeron flux factor is assumed to
have the following form
\begin{equation}
 f_{\p/p}(x_{\p},t) = x_{\p}^{1-2\alpha_ {\p}(t)} \exp (B_{\p} t).
 \label{eq:12}
\end{equation}
$\alpha _{\p}(t)$ is the pomeron trajectory, $\alpha _{\p}(t)=\alpha _{\p}
(0) + \alpha'_{\p} t$, assumed to be linear in $t$. The values of $B_{\p},
\alpha _{\p}(0)$ and $\alpha _{\p}'$ are taken from \cite{H1} and have the
values $B_{\p}=4.6$ GeV$^{-2}$, $\alpha_{\p}(0)=1.17$, and $\alpha_{\p}'=
0.26$ GeV$^{-2}$. Usually $f_{\p/p}(x_{\p},t)$ as written in Eq.\
(\ref{eq:12}) has in addition to the dependence on $x_{\p}$ and $t$ a
normalization factor $N$, which can be inferred from the asymptotic behavior
of $\sigma_{\rm tot}$  for $pp$  and $p\bar{p}$ scattering. Since it is
unclear whether these soft diffractive cross sections are dominated by a
single pomeron exchange, it is better to include $N$ into the pomeron PDFs
$f_{a/\p}$ and fix it from the diffractive DIS data \cite{H1}. The
diffractive DIS cross section $\sigma _r^{D(3)}$ is measured in the
kinematic range $6.5 \leq Q^2 \leq 120$ GeV$^2$, $0.01 \leq \beta \leq 0.9$
and $10^{-4} \leq x_{\p} < 0.05$.

The pomeron couples to quarks in terms of a light flavor singlet $\Sigma
(z_{\p})=u(z_{\p})+d(z_{\p})+s(z_{\p})+\bar{u}(z_{\p})+\bar{d}(z_{\p})+
\bar{s}(z_{\p})$ and to gluons in terms of $g(z_{\p})$, which are
parameterized at the starting scale $Q_0=\sqrt{3}$ GeV. $z_{\p}$ is the
momentum fraction entering the hard subprocess, so that for the LO process
$z_{\p}=\beta$, and in NLO $\beta < z_{\p} < 1$. These PDFs of the pomeron
are parameterized by a particular form in terms of Chebychev polynomials as
given in \cite{H1}. Charm quarks couple differently from the light quarks by
including the finite charm mass $m_c=1.5$ GeV in the massive charm scheme
and describing the coupling to photons via the photon-gluon fusion process.
For the pomeron PDFs, we used a two-dimensional fit in the variables
$z_{\p}$ and $Q^2$ and then inserted the interpolated result in the cross
section formula.

\subsection{Cross Section Formula}

Under the assumption that the cross section can be calculated from the well
known formul\ae{} for jet production in low $Q^2$ $ep$ collisions, the cross
section for the reaction $e+p \rightarrow e+2~{\rm jets}+X'+Y$ is computed
from the following basic formula:
\begin{eqnarray}
 && d\sigma ^D(ep \rightarrow e+2~{\rm jets}+X'+Y) = \nonumber \\ 
 && \sum_{a,b} \int_{t_{\rm cut}}^{t_{\min}}dt \int_{x_{\p}^{\min}}^{x_{\p}
  ^{\max}} dx_{\p} \int_{0}^{1}dz_{\p} \int_{y_{\min}}^{y_{\max}}dy \int_{0}
  ^{1}dx_{\gamma} \nonumber \\
 && f_{\gamma/e}(y) f_{a/\gamma }(x_\gamma,M^2_{\gamma }) f_{\p/p}(x_{\p},t)
  f_{b/\p}(z_{\p},M_{\p}^2) \nonumber \\
 && d\sigma^{(n)} (ab \rightarrow {\rm jets}).
 \label{eq:13}
\end{eqnarray}
$y$, $x_\gamma$ and $z_{\p}$ denote the longitudinal momentum fractions of
the photon in the electron, the parton $a$ in the photon, and the parton $b$
in the pomeron. $M_{\gamma}$ and $M_{\p}$ are the factorization scales at
the respective vertices, and $d\sigma ^{(n)}(ab \rightarrow {\rm jets)}$ is
the cross section for the production of an $n$-parton final state from two
initial partons $a$ and $b$. It is calculated in LO and NLO, as are the
PDFs of the photon and the pomeron.

The function $f_{\gamma /e}(y)$, which describes the virtual photon
spectrum, is assumed to be given by the well-known Weizs\"acker--Williams
approximation,
 \begin{eqnarray}
 f_{\gamma /e}(y) &=& \frac{\alpha }{2\pi} \left[\frac{1+(1-y)^2}{y} \ln 
\frac{Q^2_{\max}(1-y)}{m_e^2y^2} \right. \nonumber \\
 &+& \left. 2m_e^2y(\frac{1-y}{m_e^2y^2} -\frac{1}{Q^2_{\max}})\right].
\end{eqnarray}
Usually, only the dominant leading logarithmic contribution is considered.
We have added the second non-logarith\-mic term as evaluated in \cite{Fri}.
$Q^2_{\max} = 0.01$ GeV$^2$ for the cross sections calculated in this work.

The formula for the cross section $d\sigma ^D$ can be used for the resolved
as well as for the direct process. For the latter, the parton $a$ is the
photon and $f_{\gamma /\gamma }(x_\gamma,M^2_{\gamma})=\delta(1-x_\gamma)$,
which does not depend on $M_{\gamma }$. As is well known, the distinction
between direct and resolved photon processes is meaningful only in LO of
perturbation theory. In NLO, collinear singularities arise from the photon
initial state, that must be absorbed into the photon PDFs and produce a
factorization scheme dependence as in the proton and pomeron cases. The
separation between the direct and resolved processes is an artifact of
finite order perturbation theory and depends in NLO on the factorization
scheme and scale $M_{\gamma}$. The sum of both parts is the only physically
relevant quantity, which is approximately independent of the factorization
scale $M_{\gamma}$ due to the compensation of the scale dependence between
the NLO direct and the LO resolved contribution \cite{BKS,KKK}.

For the resolved process, PDFs of the photon are need\-ed, for which we
choose the LO and NLO versions of GRV \cite{GRV}. They have been found to
give a very good description of the cross sections for photoproduction of
inclusive one- and two-jet final states \cite{H12,ZEUS}.

\section{Results}
\label{sec:3}

In this Section, we present the comparison of the theoretical predictions in
LO and NLO with the experimental data from H1 \cite{H13}. In this paper,
preliminary data on cross sections differential in $x_{\gamma}^{\rm jets}$
and $z_{\p}^{\rm jets}$ for the diffractive production of two jets in the
kinematic regions specified in Tab.~\ref{tab:1} are given. These two cross
sections are the only differential cross sections, which are not normalized
to unity in the measured kinematic range. All other differential cross
sections, namely those differential in the variables $\log_{10}x_{\p}$, $y$,
$E_T^{\rm jet1}$, $M_X^{\rm jets}$, $M_{12}^{\rm jets}$, $\overline{\eta}
^{\rm jets}$, and $|\Delta \eta^{\rm jets}|$, are normalized cross sections.
With these latter distributions, only the shape can be used to test a
possible factorization breaking in the resolved component.

\begin{figure*}
 \centering
 \epsfig{file=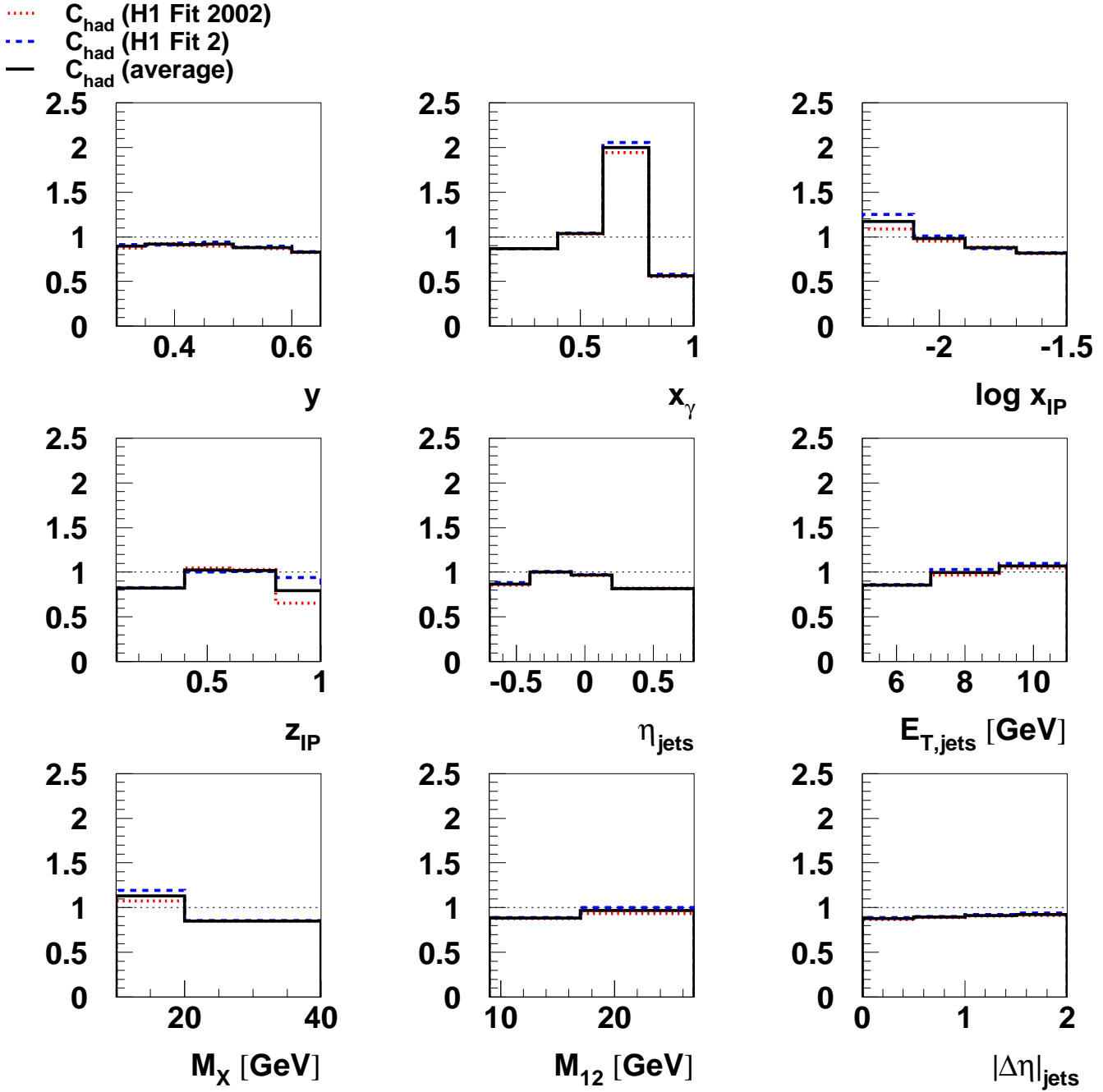,bbllx=24,bblly=154,bburx=576,bbury=695,%
         width=\textwidth}
 \caption{\label{fig:2}Ratios of hadronic to partonic dijet cross sections
 for two different fits of the parton densities in the pomeron \cite{H1,H11}
 as well as their average \cite{Schi}.}
\end{figure*}

Before we confronted the calculated cross sections with the experimental
data, we have corrected them for hadroni\-zation effects. The calculated
cross sections are the cross sections for the production of QCD jets, which
consist either of one parton or a recombination of two partons according to
the $k_T$-cluster algorithm. The experimental cross sections are measured
with hadron jets constructed with the same jet algorithm. Although the
difference between the two kinds of jets is not large, in particular for
jets with sufficiently large $E_T$'s, we have corrected the originally
calculated cross sections with a factor $C_{\rm had}$ for the transformation
from QCD jets to hadron jets. The correction factors $C_{\rm had}$ for the
differential cross sections in the kinematic variables of interest are shown
in Fig.~\ref{fig:2}.
%
%
Here, $C_{\rm had}$ is the ratio of the respective cross sections for
hadronic jets to partonic jets. The correction factors have been calculated
from Monte Carlo models including LO cross sections together with parton
showering and subsequent hadronization by the H1 group \cite{Schi}. As seen
in Fig.~\ref{fig:2}, $C_{\rm had}$ is approximately equal to one with
deviations less than $20\%$. The only exception is $C_{\rm had}$ for the
$x_{\gamma}^{\rm jets}$ cross section with a value that is appreciably
different from one for $x_{\gamma}^{\rm jets} \geq 0.6$.

The differential cross sections have been calculated in LO and NLO with
varying scales, where the renormalization scale and both factorization
scales are set equal and are $\mu = \xi E_T^{\rm jet1}$ with $\xi$ varied in
the range $0.5 \leq \xi \leq 2$. This way we hope to have a reasonable
estimate of the error for the theoretical cross sections and are not in
danger to base our conclusions concerning factorization breaking only on one
particular scale choice. Note that for the pomeron PDFs the variation of the
factorization scale is restricted by their parameterization to $M_{\p}^2\leq
150$ GeV$^2$.

The theoretical cross sections are presented in two versions in LO and NLO,
respectively. In the first version no suppression factor $R$ is applied. It
corresponds to the LO or NLO prediction with no factorization breaking, 
labeled $R=1$ in the figures. The second version is with a suppression
factor $R=0.34$ in the resolved cross section, labelled $R=0.34$ in the
figures. This particular value for $R$ is motivated by the recent work of 
Kaidalov et al.\ \cite{KKMR}. These authors studied the ratio of 
diffractive to inclusive dijet photoproduction in the HERA regime with and
without including unitarity effects, which are responsible for factorization
breaking, as a function of $x_{\gamma}$. In this study they applied a very
simplified dijet production model for this ratio, which is very similar
to the model proposed by the CDF collaboration for $p\bar{p}$ collisions
\cite{CDF}. From the calculations of this ratio, with and without unitarity
corrections, they obtained the suppression factor $R=0.34$ for $x_{\gamma}
\leq 0.3$ (see Fig.~6 in Ref.\ \cite{KKMR}), which they attribute to 
the resolved part of the photoproduction cross section. We shall use this
value of the suppression factor as a first try and apply it to the total
resolved part in the LO calculation and to its NLO correction. The direct
part is, in both cases, left unsuppressed ($R=1$). It is clear that not all
of the distributions will be sensitive to the value of $R$. Furthermore,
most of the distributions are normalized to one, so that the absolute
magnitude can not be used as a discriminator for the occurence of a
suppression factor.

\begin{figure*}
 \centering
 \epsfig{file=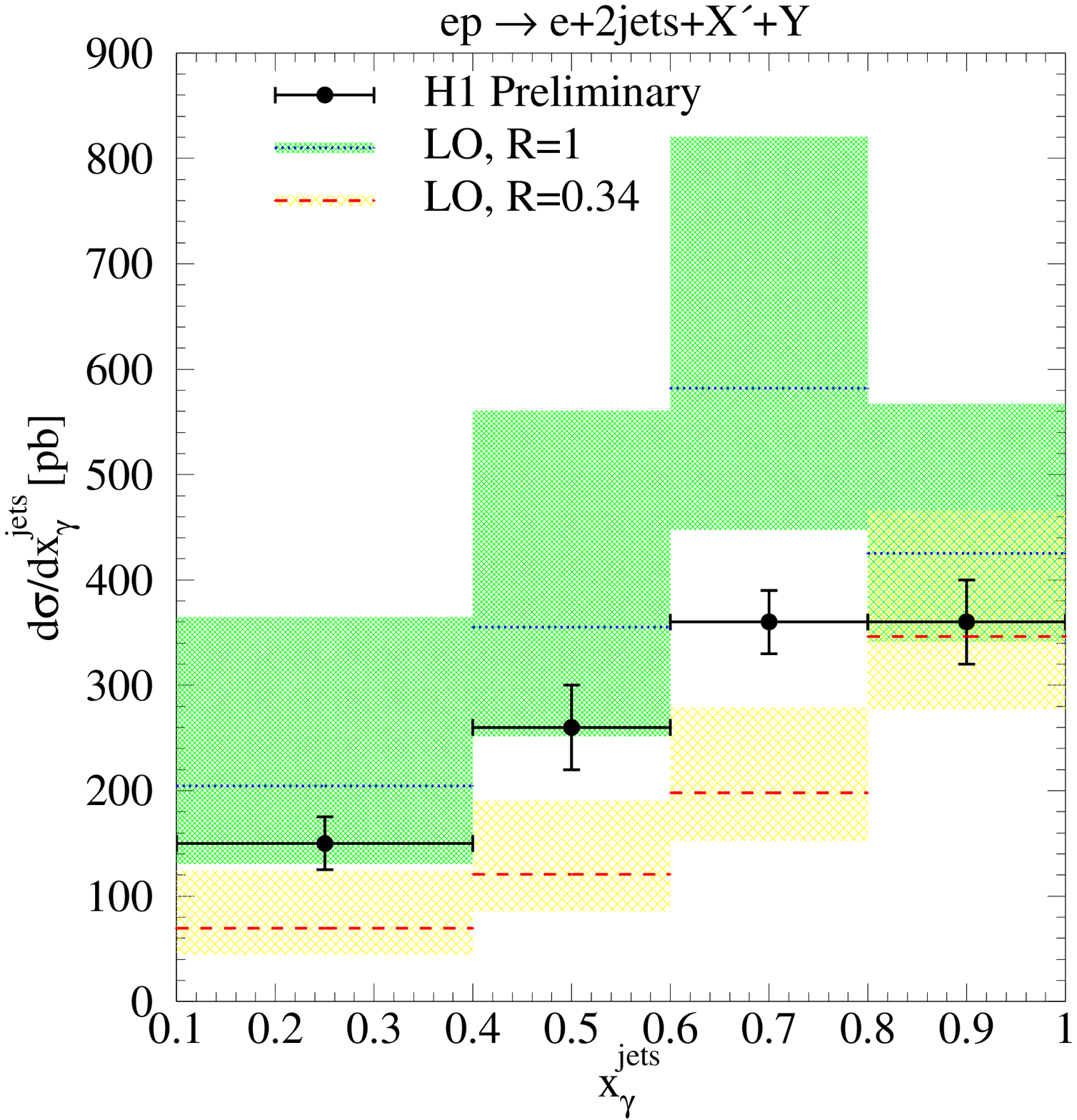,width=0.49\textwidth}
 \epsfig{file=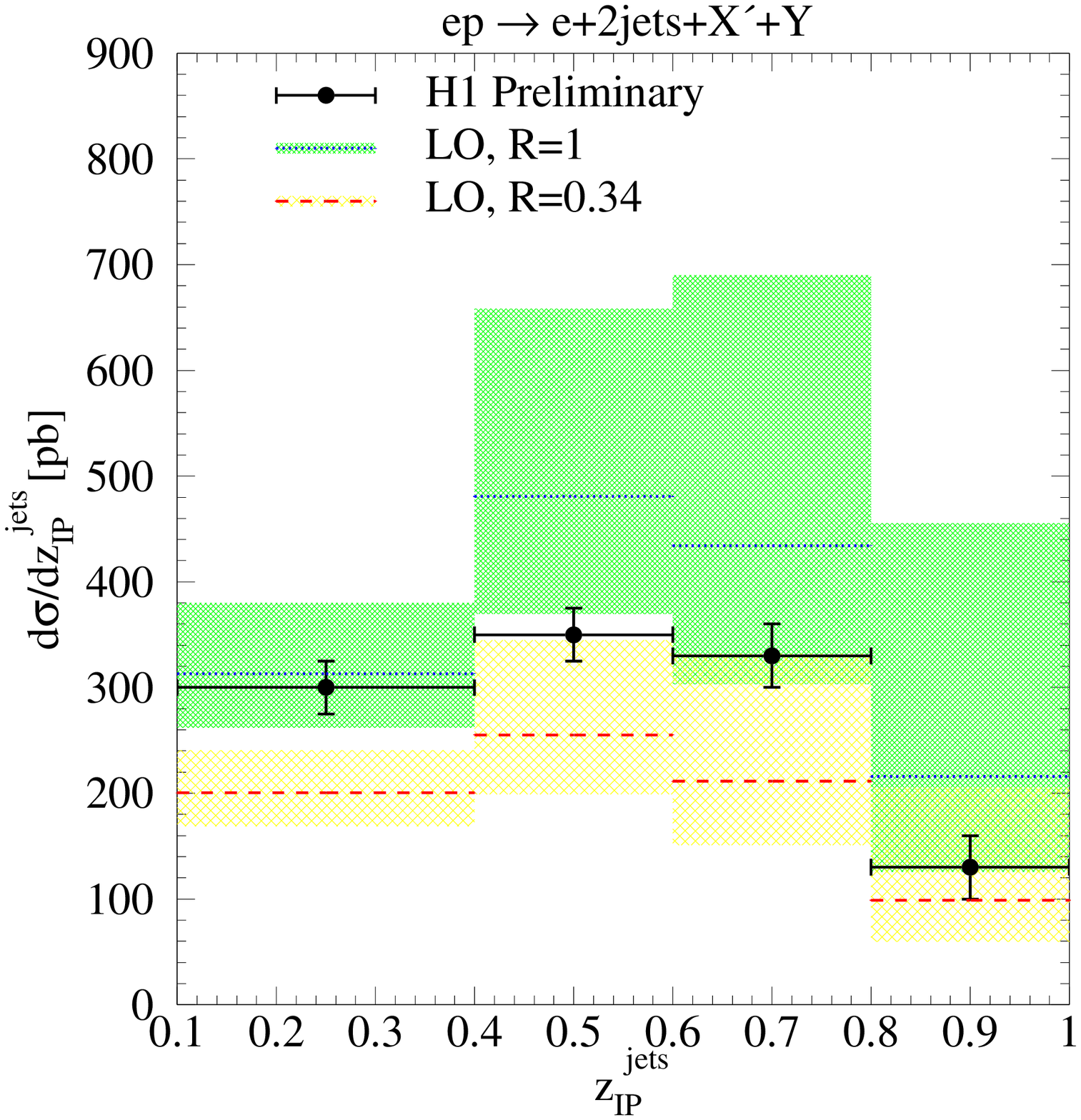,width=0.49\textwidth}
 \epsfig{file=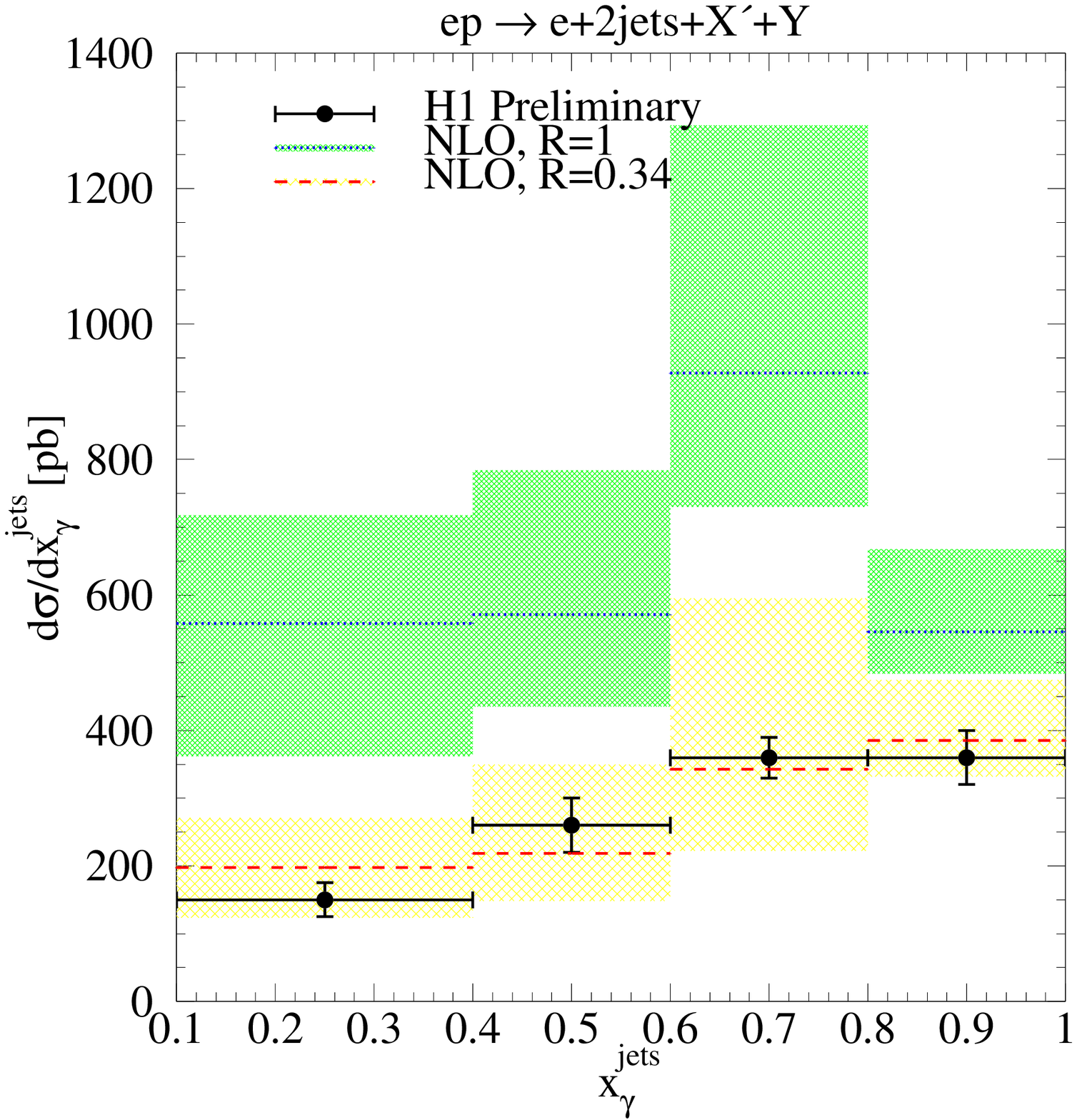,width=0.49\textwidth}
 \epsfig{file=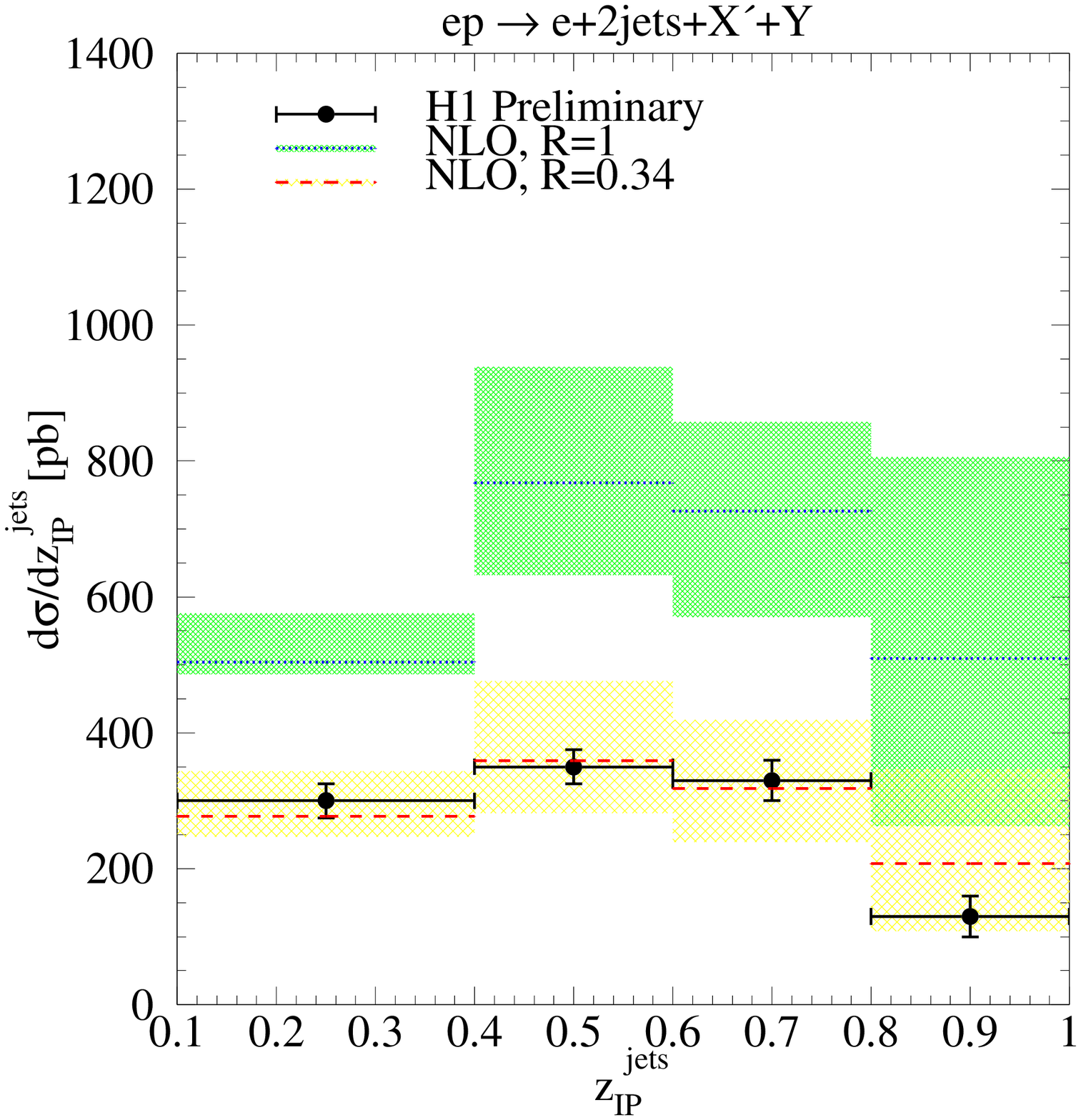,width=0.49\textwidth}
 \caption{\label{fig:3}LO (upper) and NLO (lower) cross sections for
 diffractive dijet photoproduction as functions of $x_\gamma^{\rm jets}$
 (left) and $z_{\p}^{\rm jets}$ (right), compared to preliminary H1 data.
 The shaded areas indicate a variation of scales by a factor of two around
 $E_T^{\rm jet1}$.}
\end{figure*}

\begin{figure*}
 \centering
 \epsfig{file=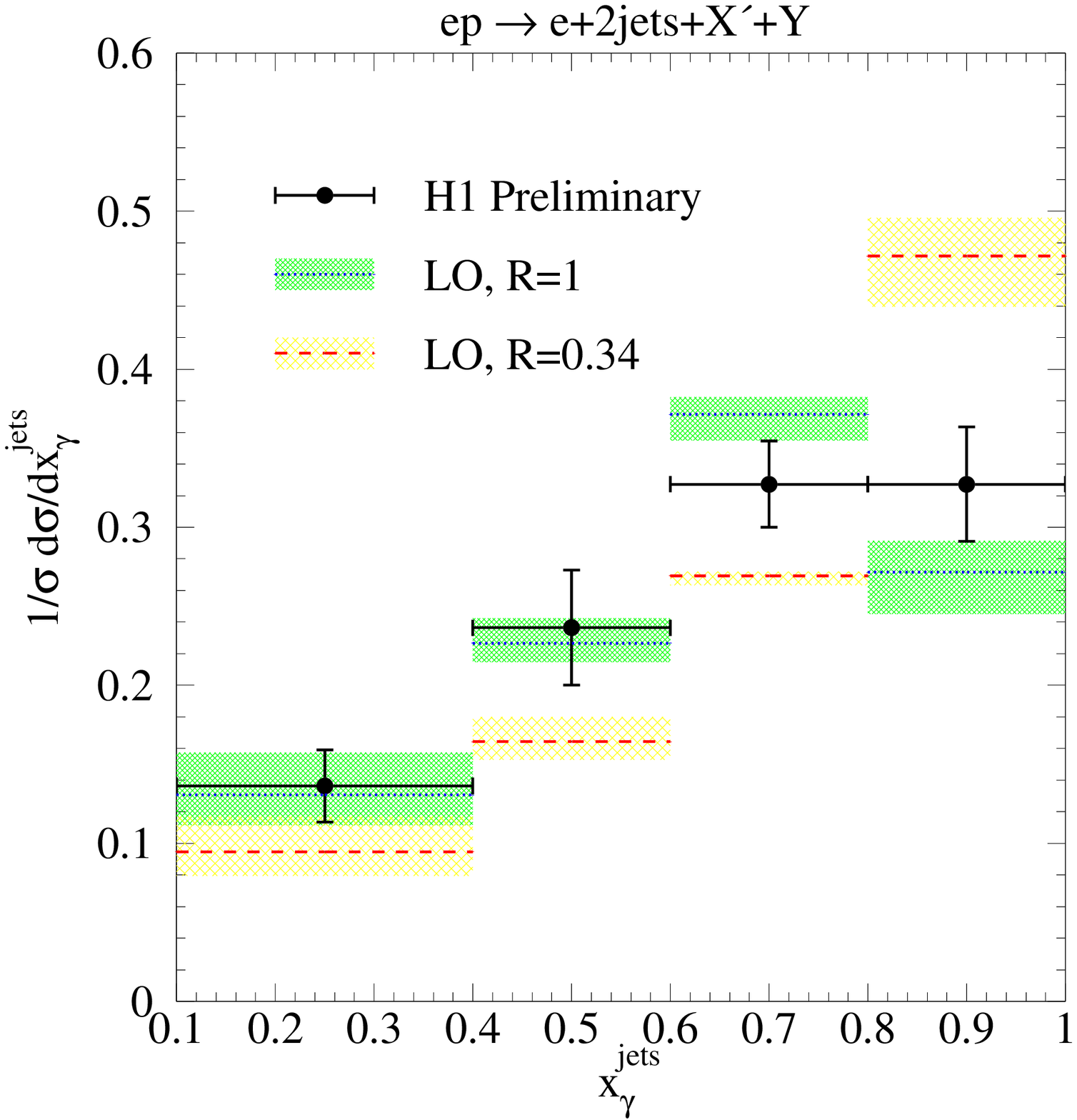,width=0.49\textwidth}
 \epsfig{file=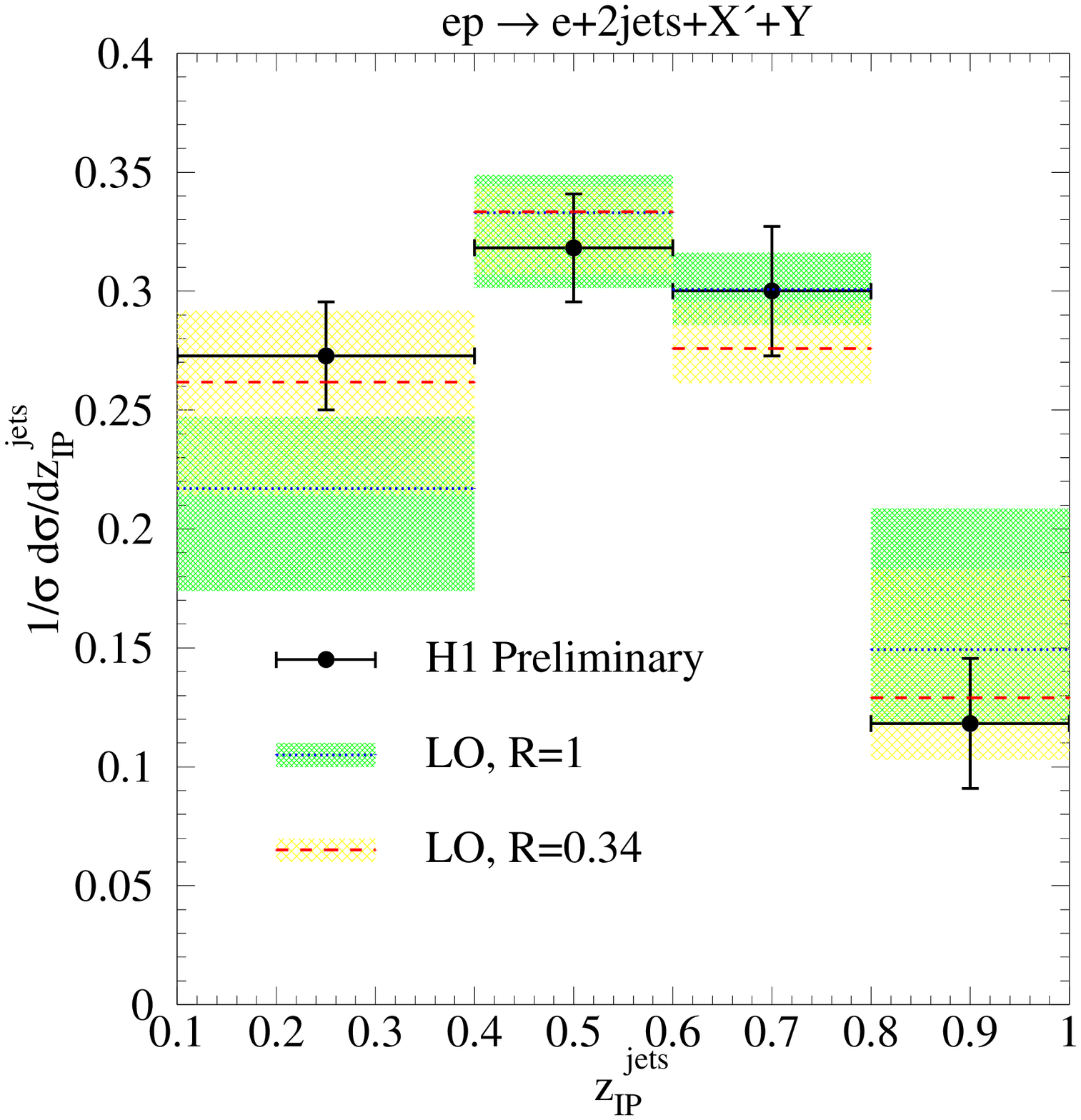,width=0.49\textwidth}
 \epsfig{file=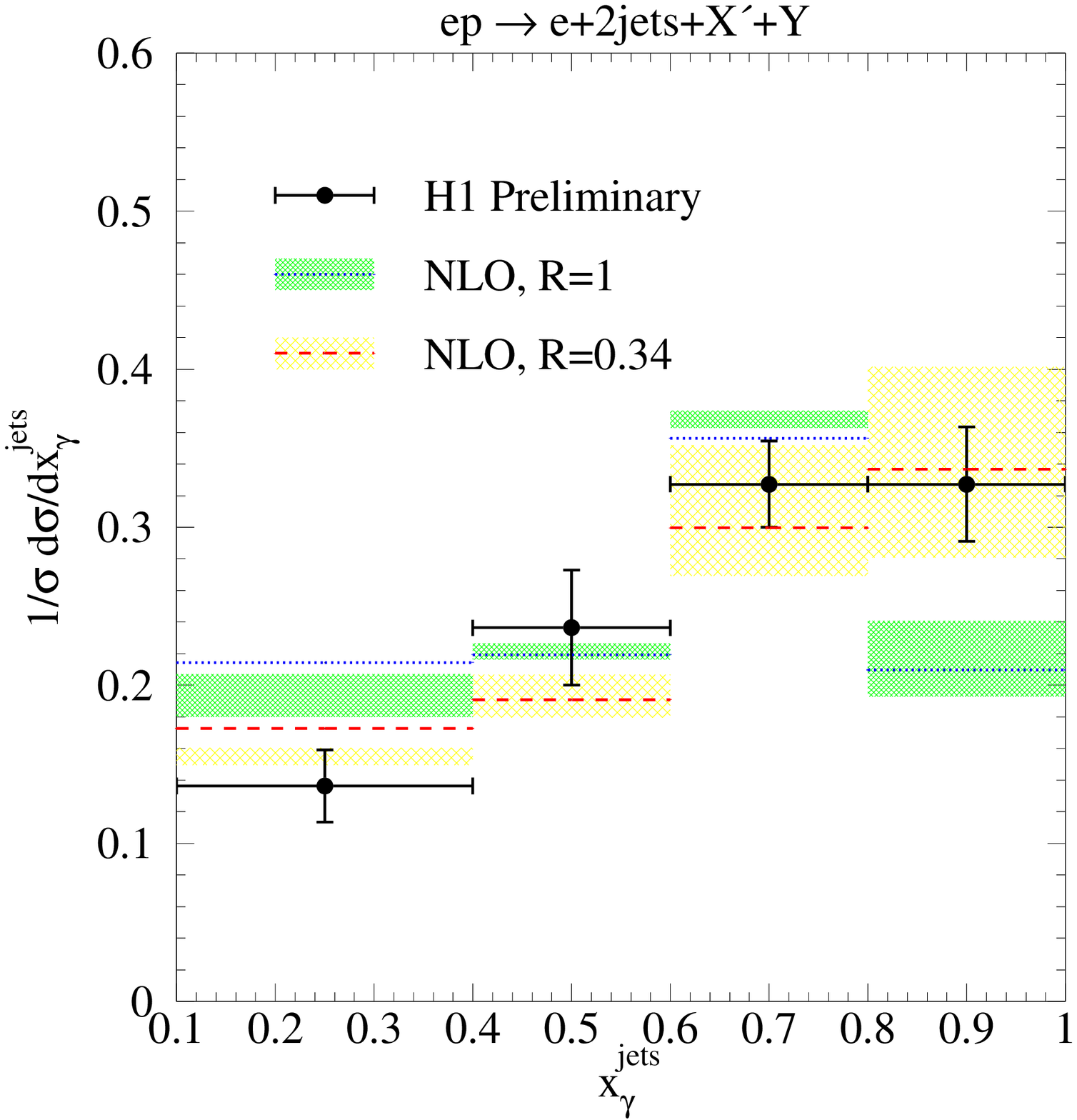,width=0.49\textwidth}
 \epsfig{file=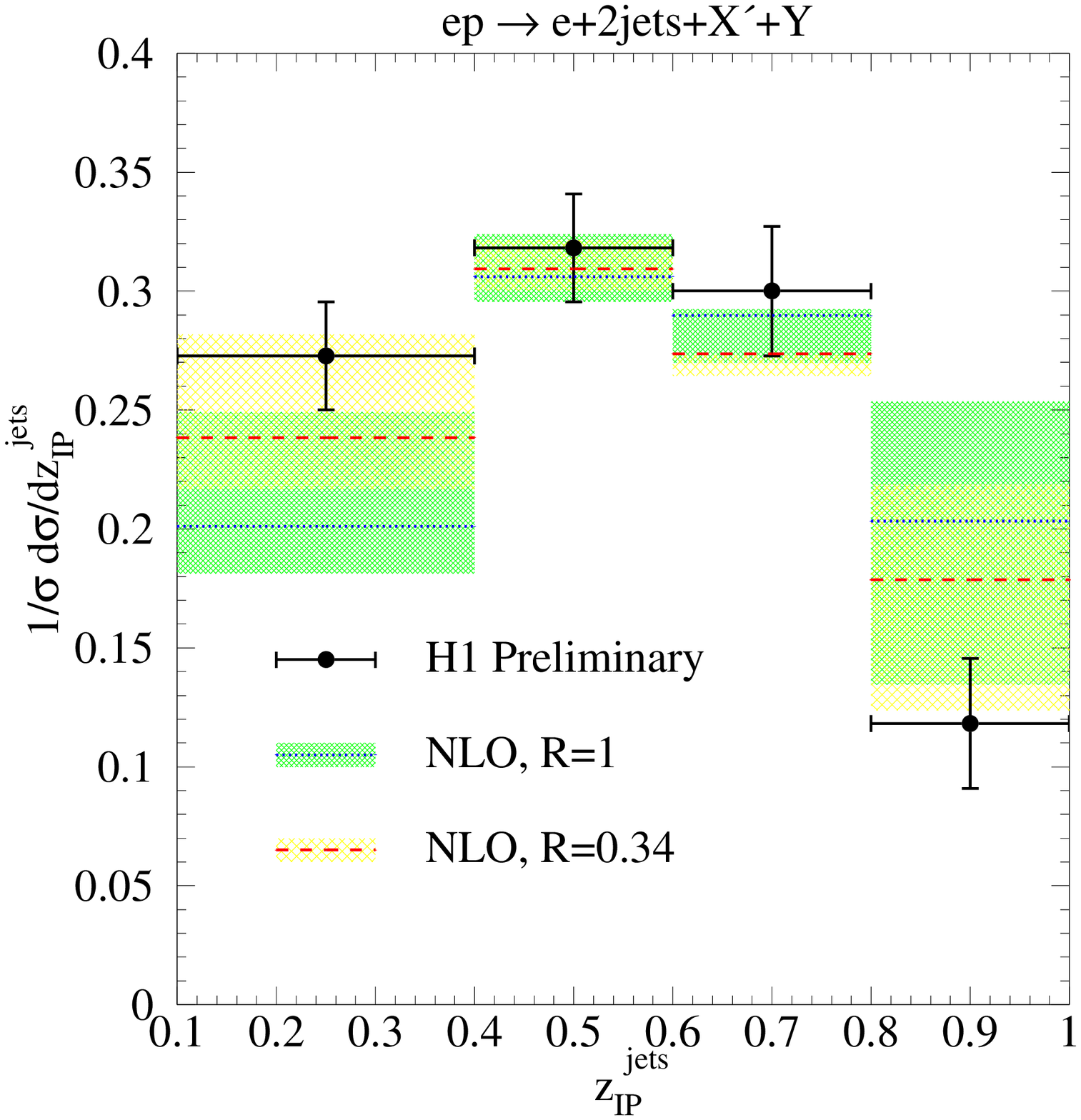,width=0.49\textwidth}
 \caption{\label{fig:3a}Normalized $x_\gamma^{\rm jets}$ (left) and $z_{\p}
 ^{\rm jets}$ (right) distributions in LO (top) and NLO (bottom), compared
 to preliminary H1 data.}
\end{figure*}

\begin{figure*}
 \centering
 \epsfig{file=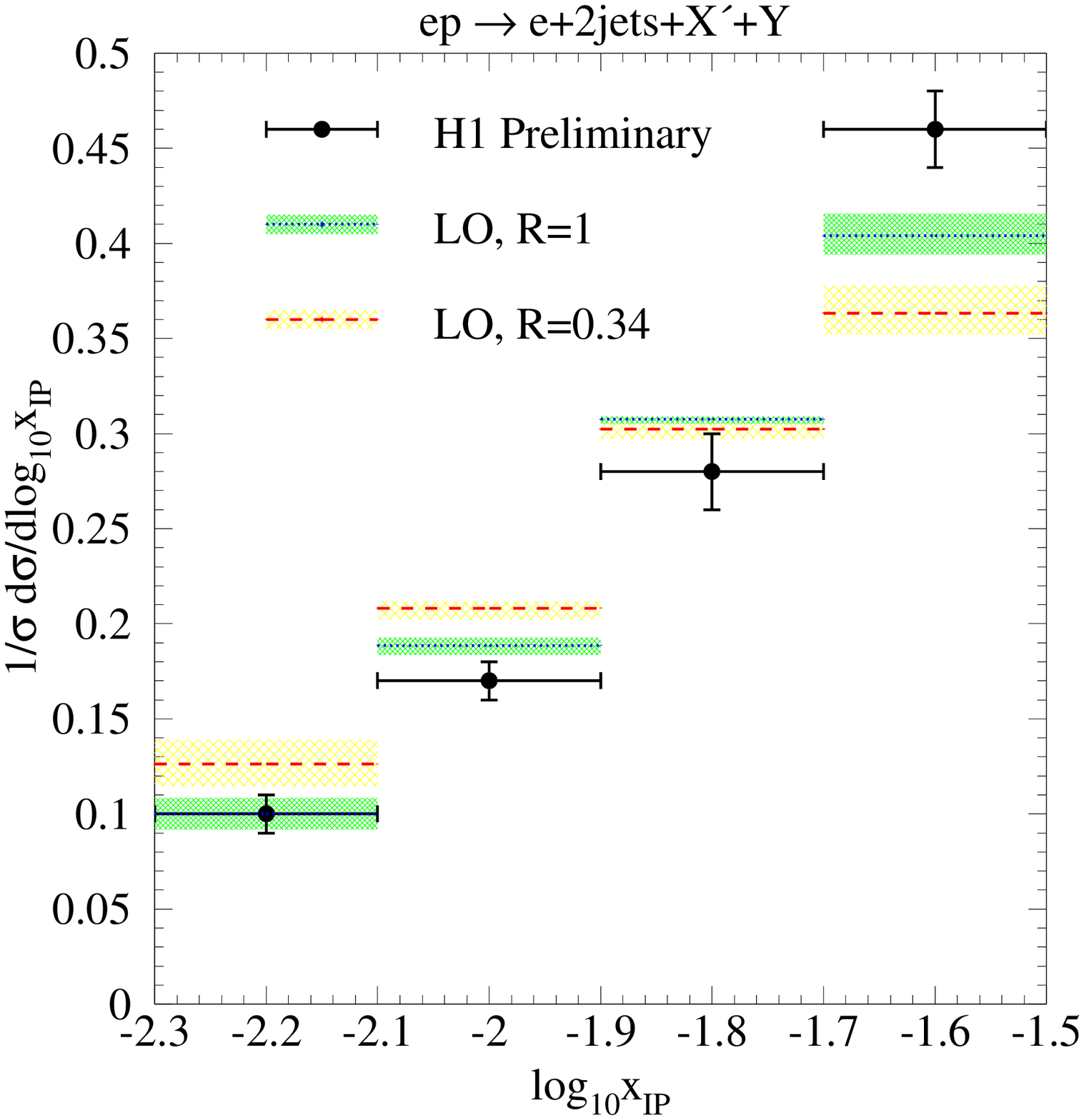,width=0.49\textwidth}
 \epsfig{file=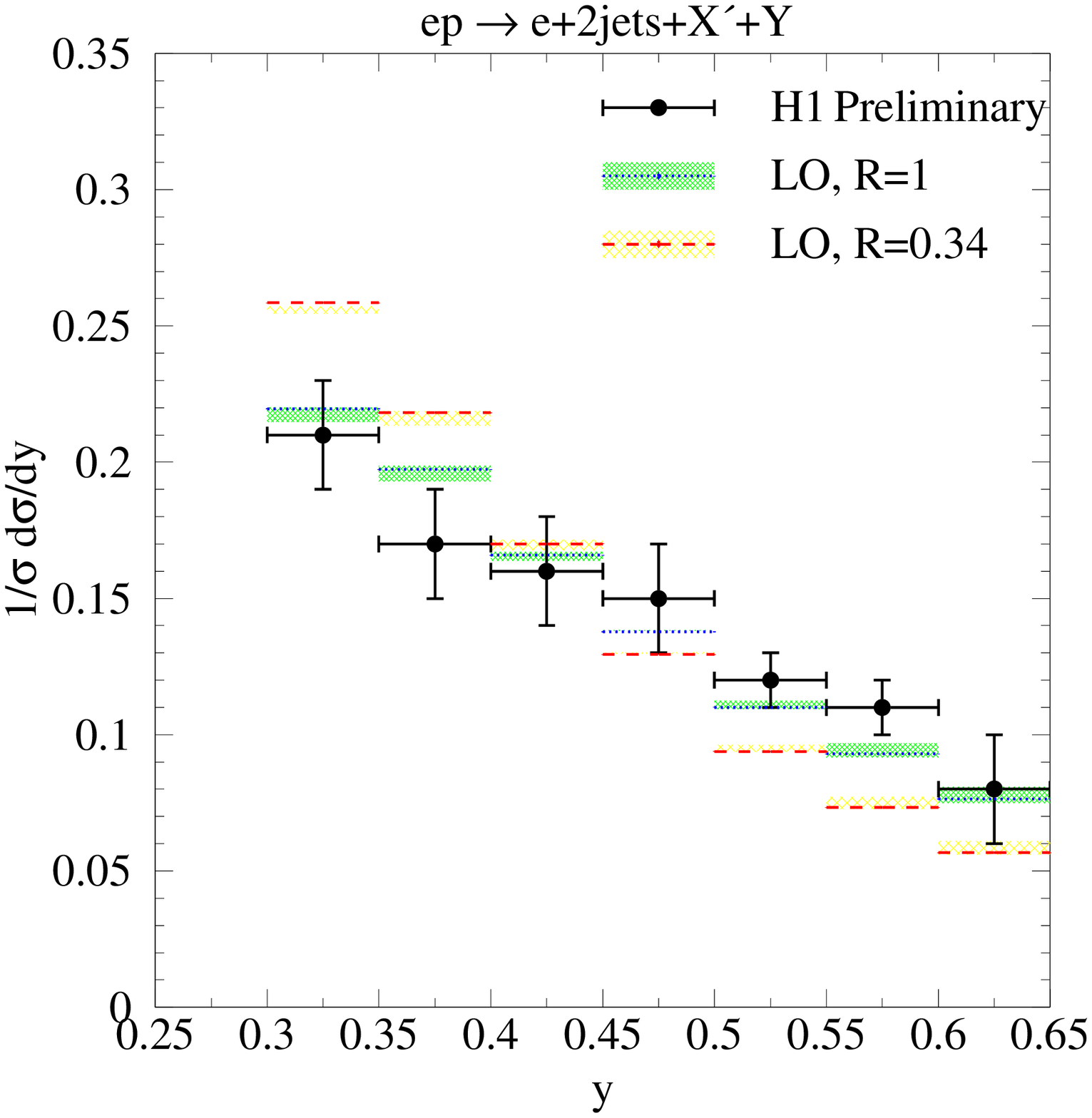,width=0.49\textwidth}
 \epsfig{file=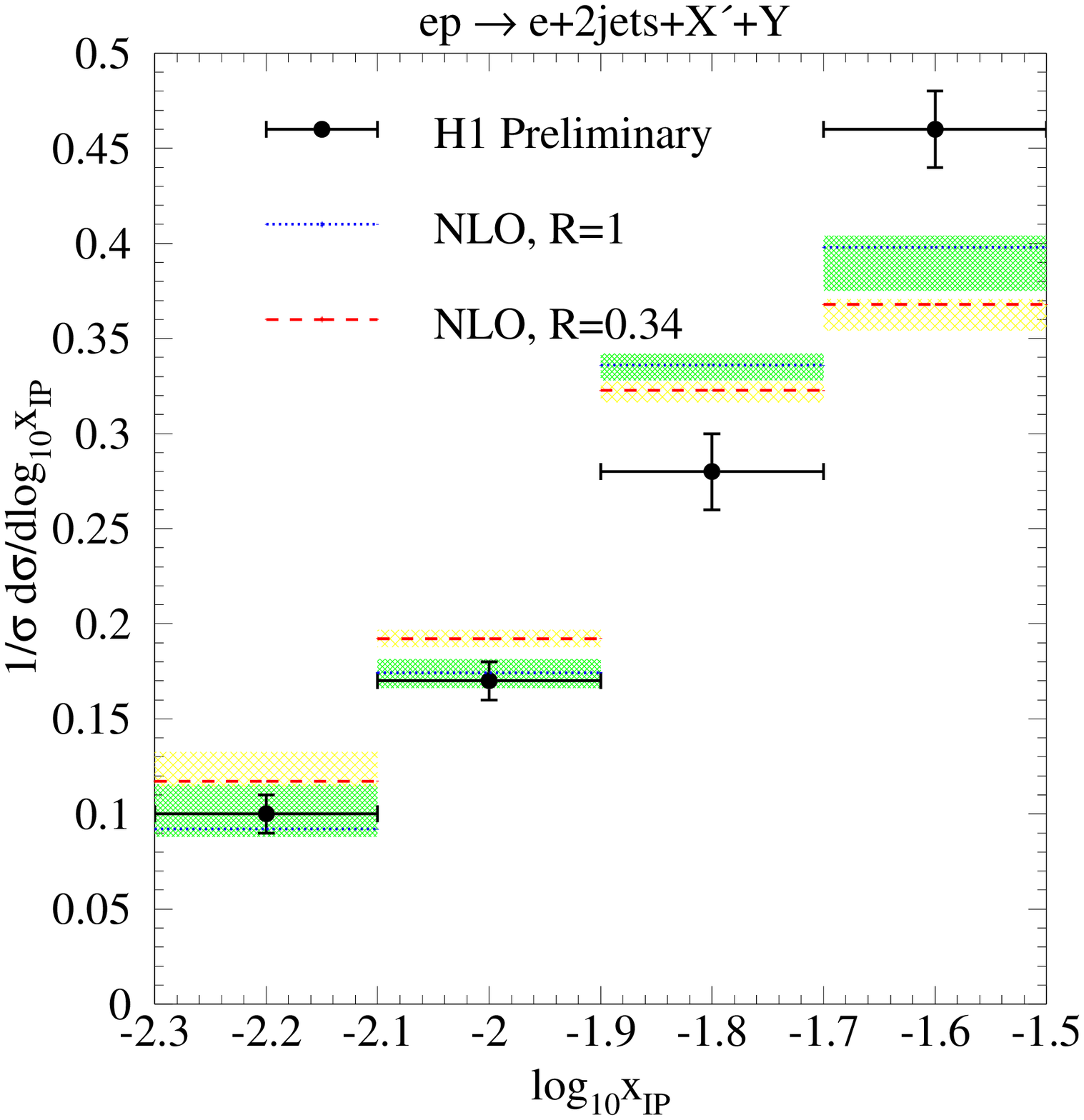,width=0.49\textwidth}
 \epsfig{file=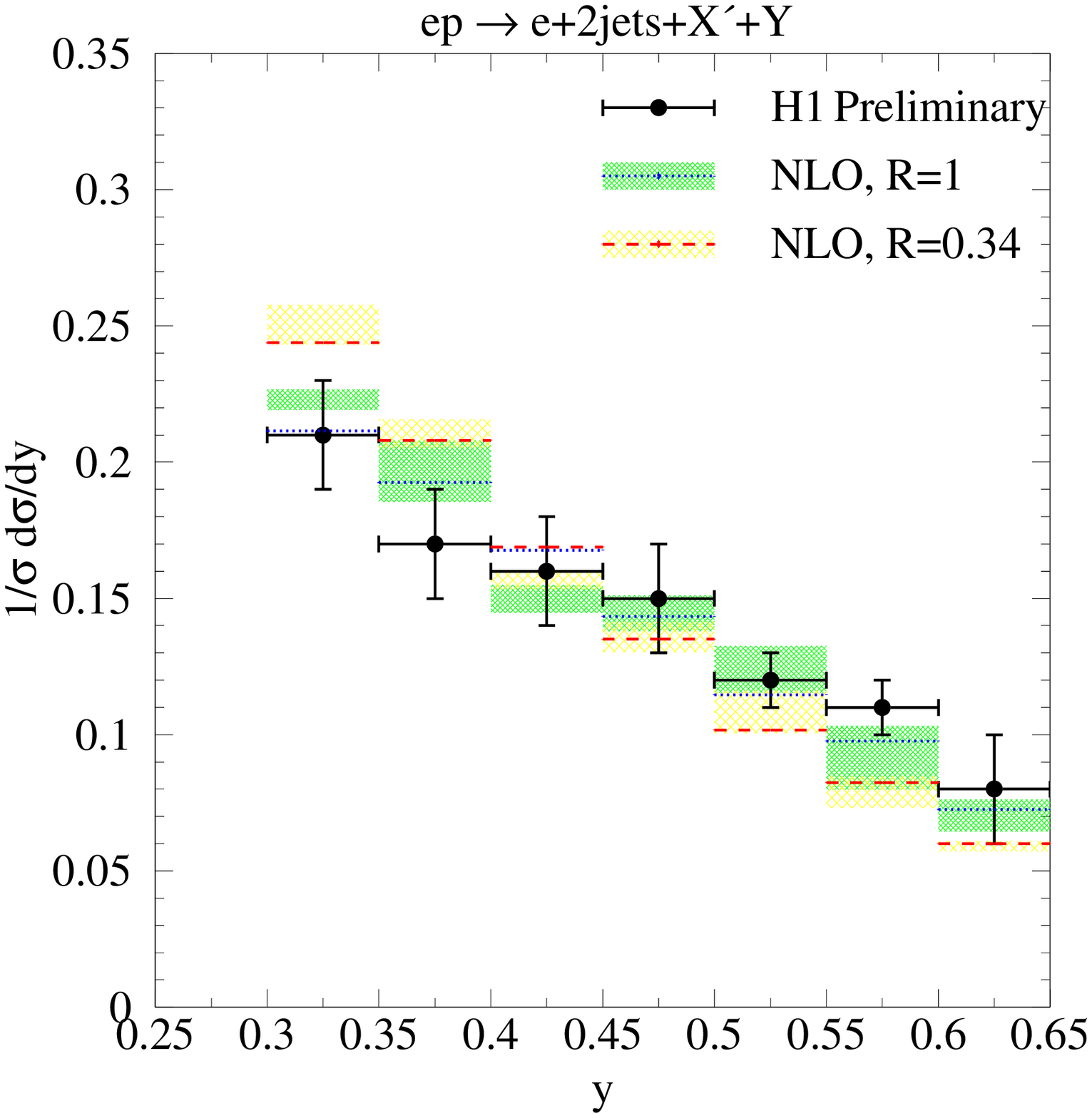,width=0.49\textwidth}
 \caption{\label{fig:4}Normalized $\log_{10}x_{\p}$ (left) and $y$ (right)
 distributions in LO (top) and NLO (bottom), compared to preliminary H1
 data.}
\end{figure*}

\begin{figure*}
 \centering
 \epsfig{file=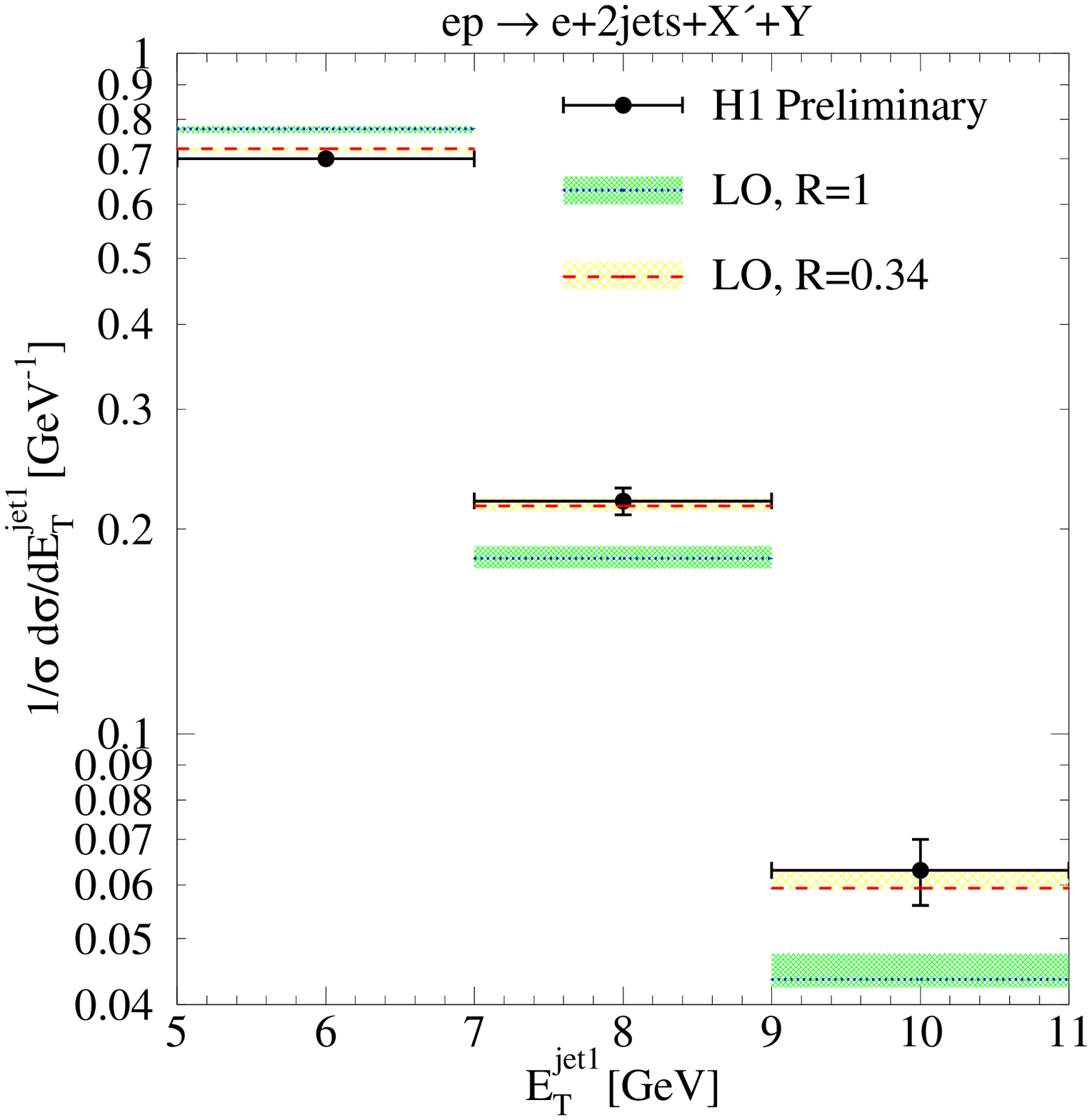,width=0.49\textwidth}
 \epsfig{file=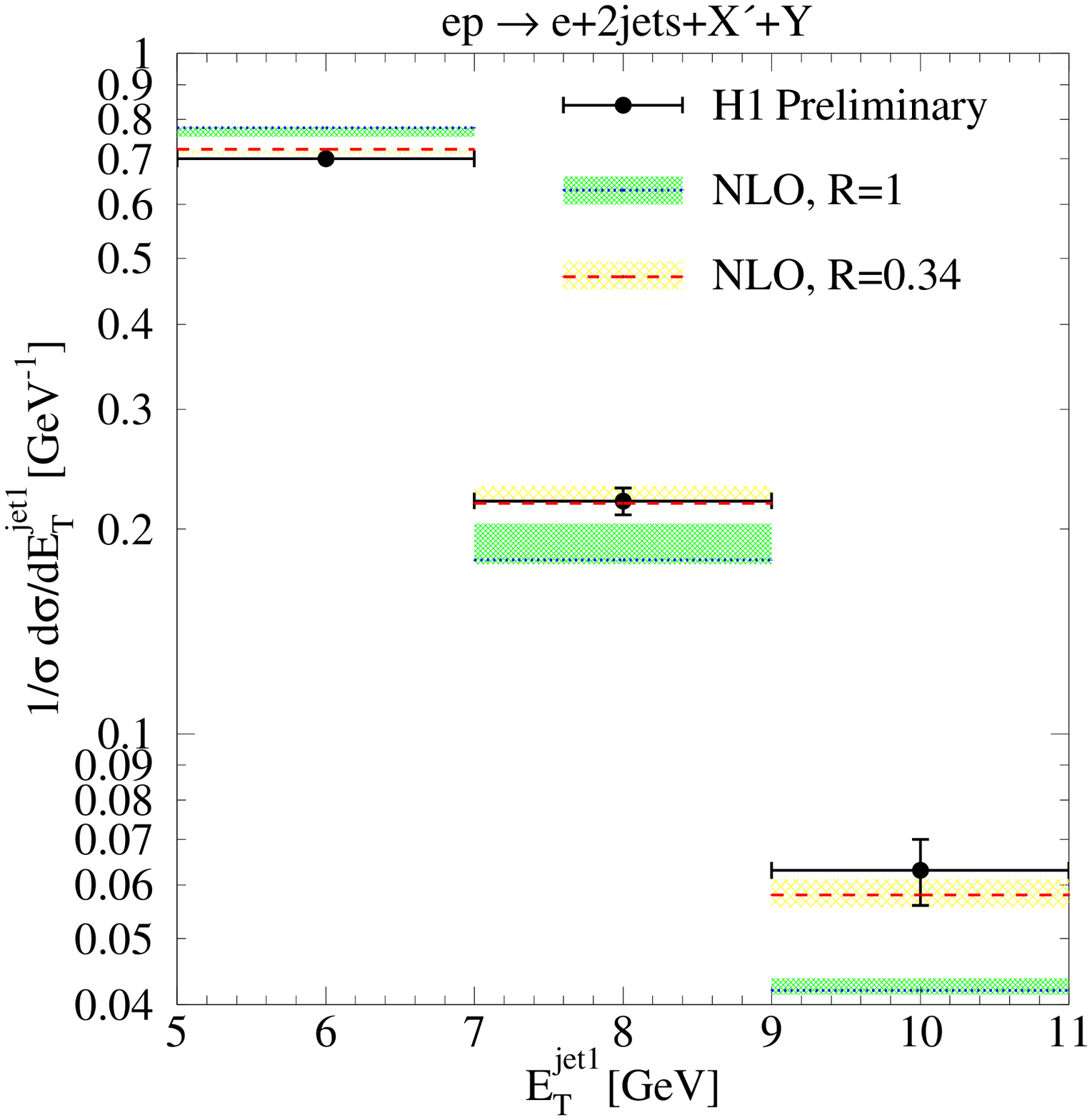,width=0.49\textwidth}
 \caption{\label{fig:5}Normalized $E_T^{\rm jet1}$ distribution in LO
 (left) and NLO (right), compared to preliminary H1 data.}
\end{figure*}

\begin{figure*}
 \centering
 \epsfig{file=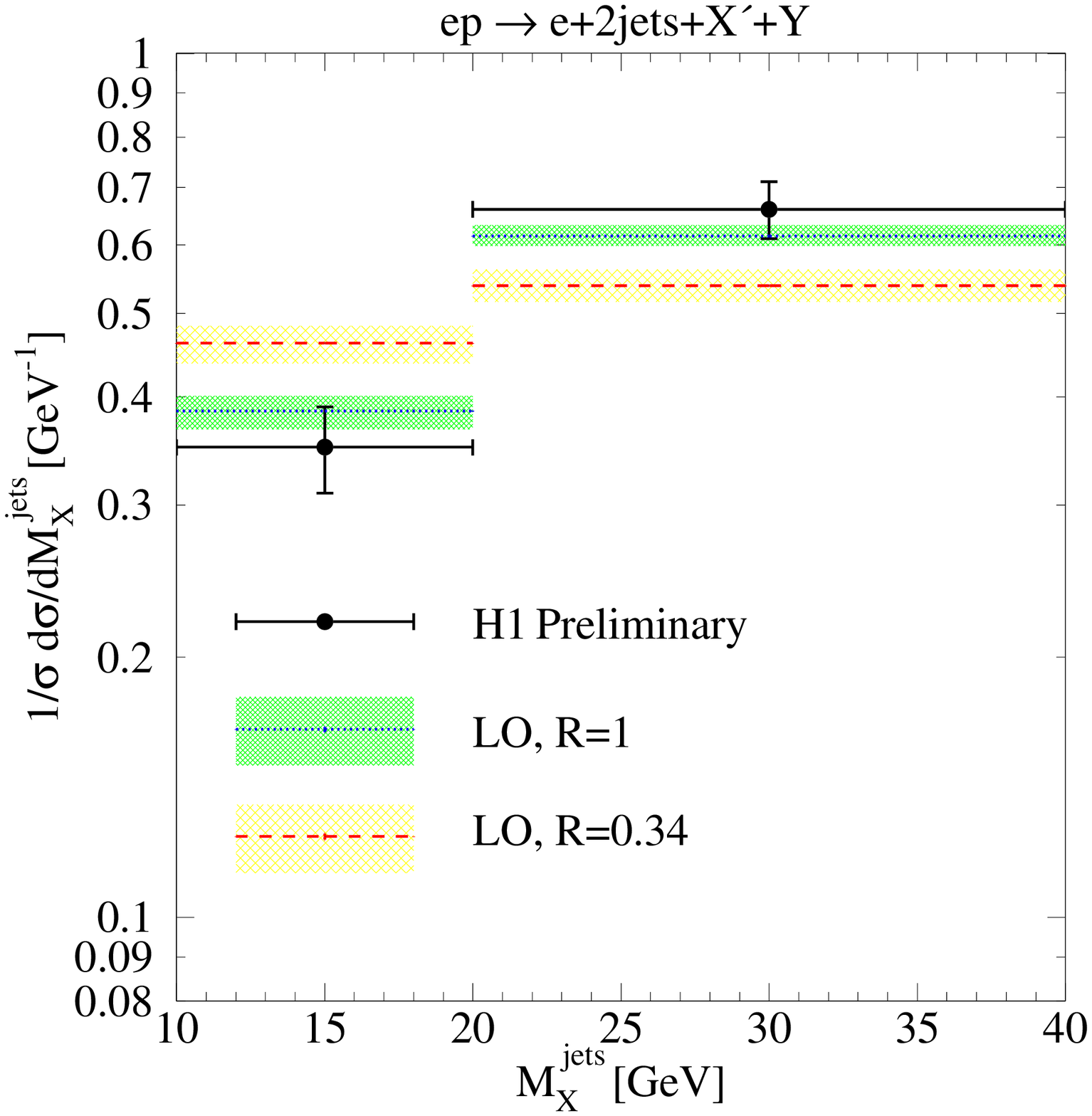,width=0.49\textwidth}
 \epsfig{file=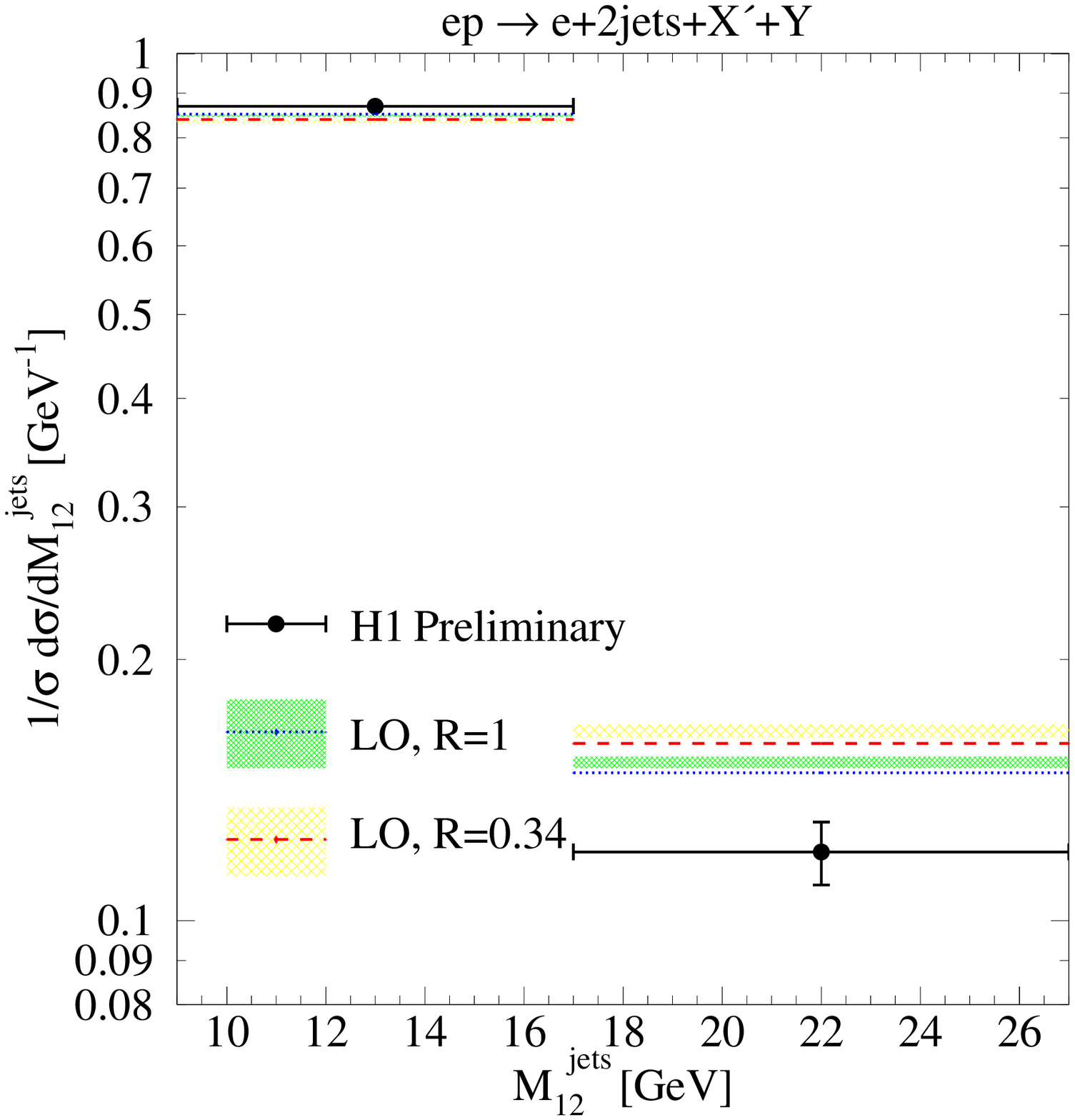,width=0.49\textwidth}
 \epsfig{file=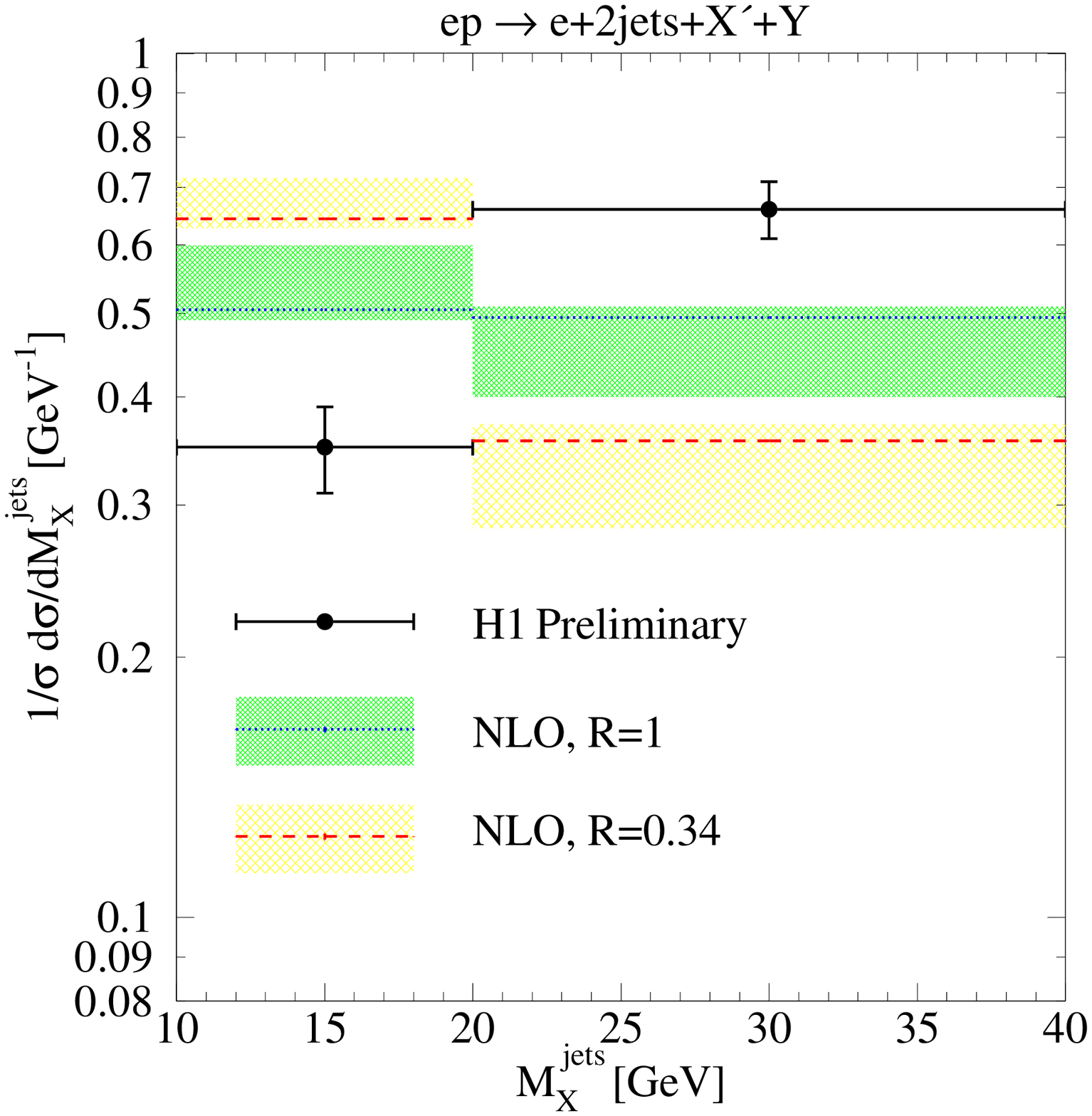,width=0.49\textwidth}
 \epsfig{file=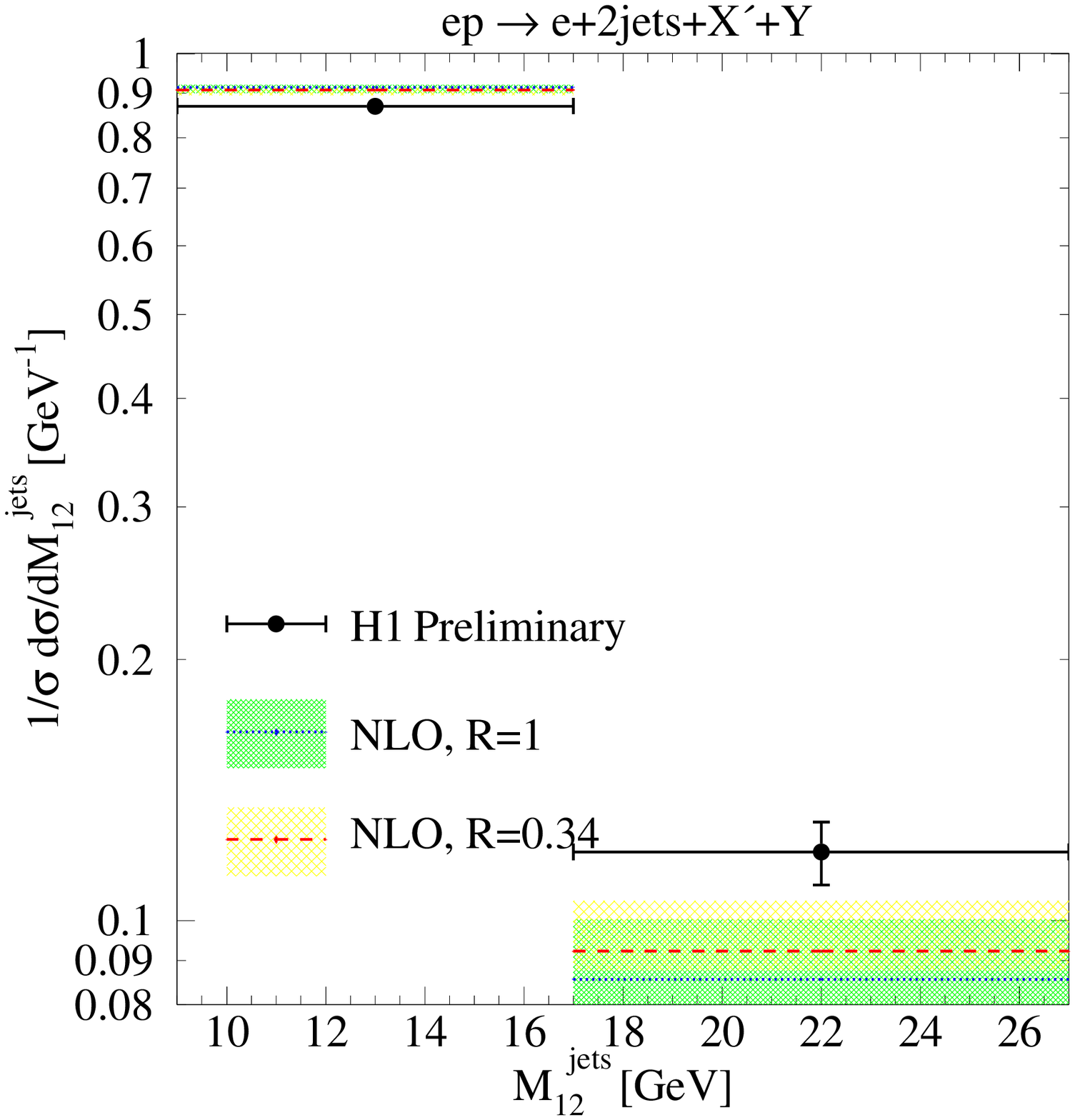,width=0.49\textwidth}
 \caption{\label{fig:6}Normalized $M_X^{\rm jets}$ (left) and $M_{12}
 ^{\rm jets}$ (right) distributions in LO (top) and NLO (bottom), compared
 to preliminary H1 data.}
\end{figure*}

\begin{figure*}
 \centering
 \epsfig{file=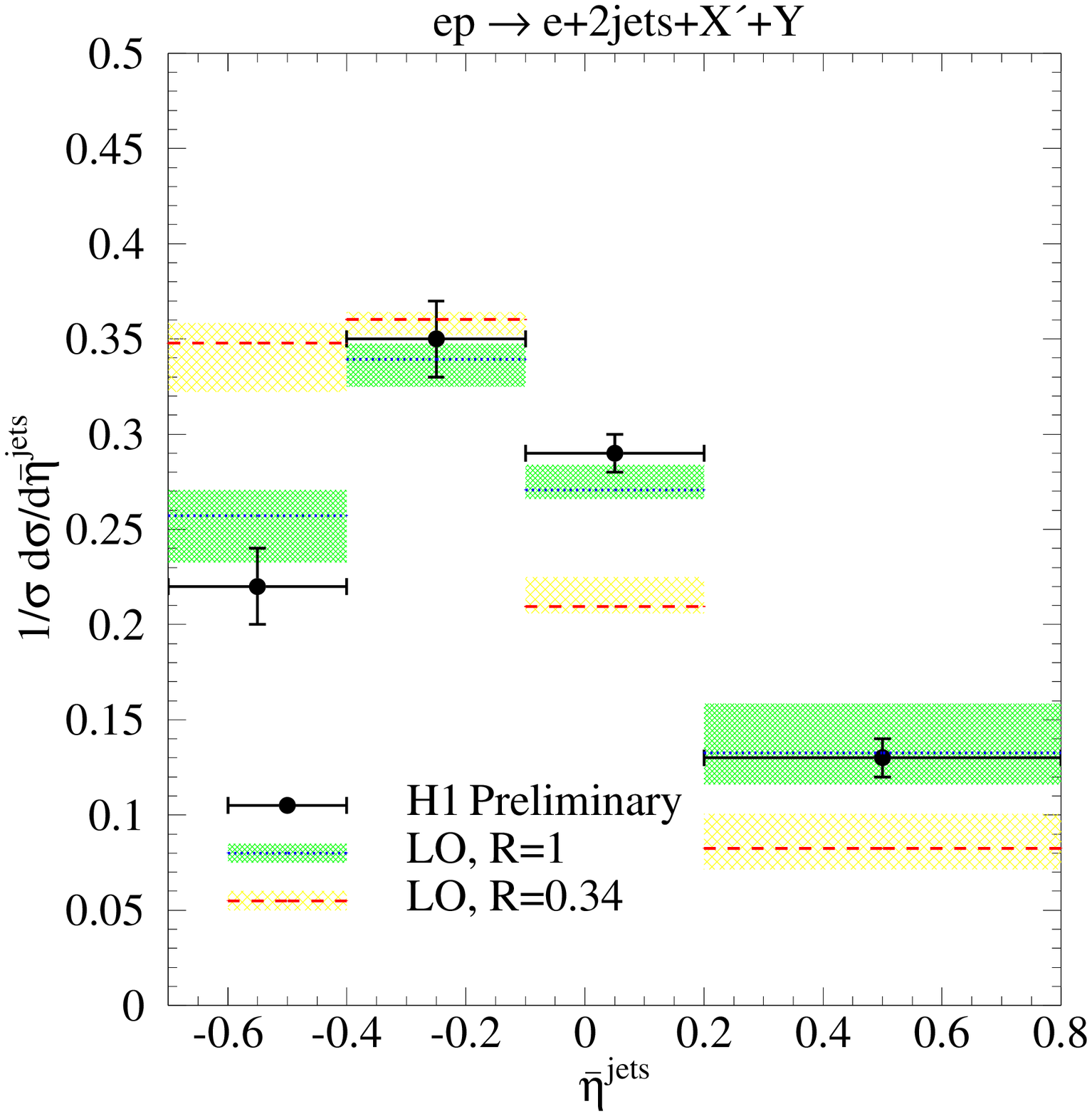,width=0.49\textwidth}
 \epsfig{file=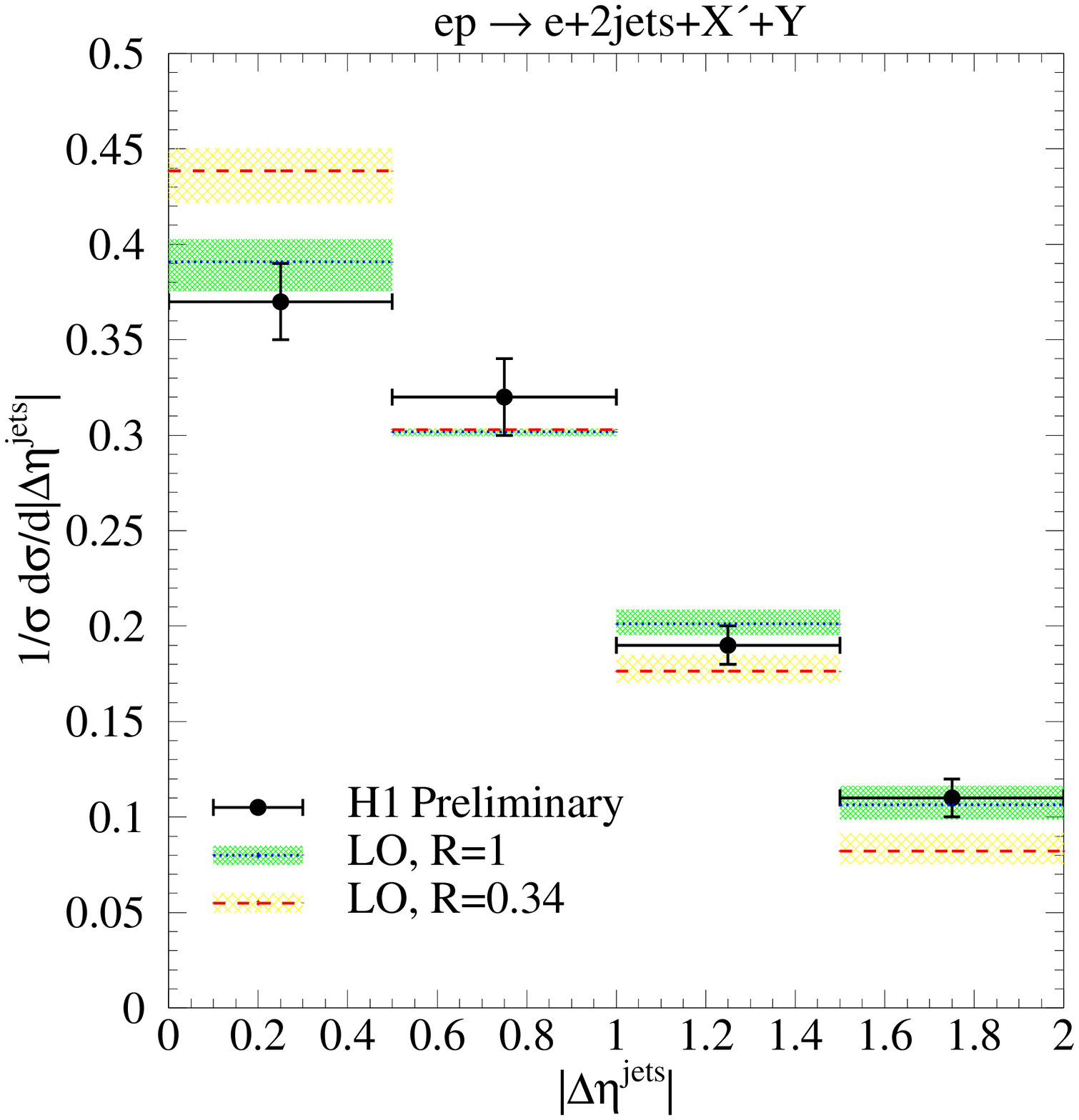,width=0.49\textwidth}
 \epsfig{file=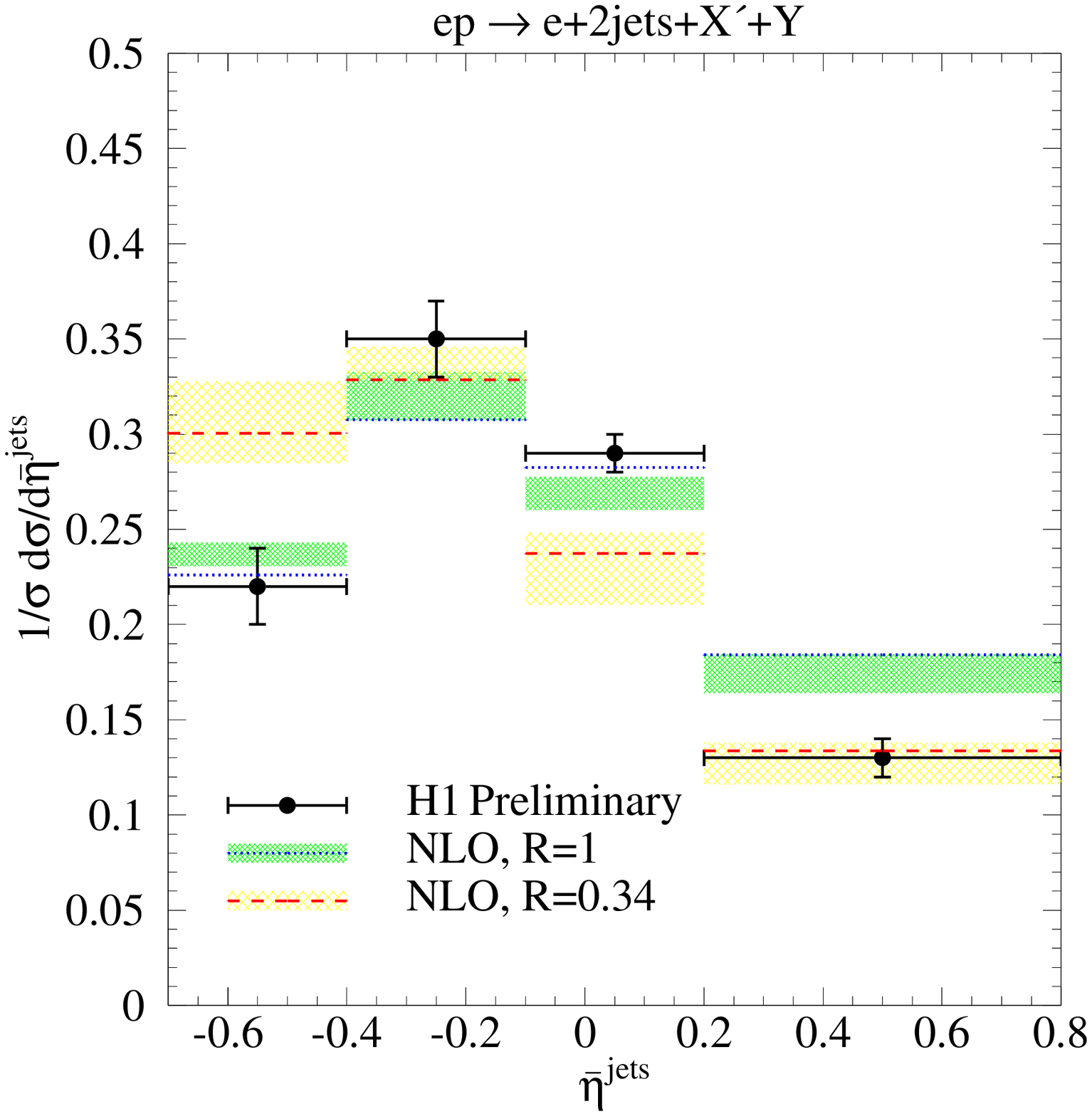,width=0.49\textwidth}
 \epsfig{file=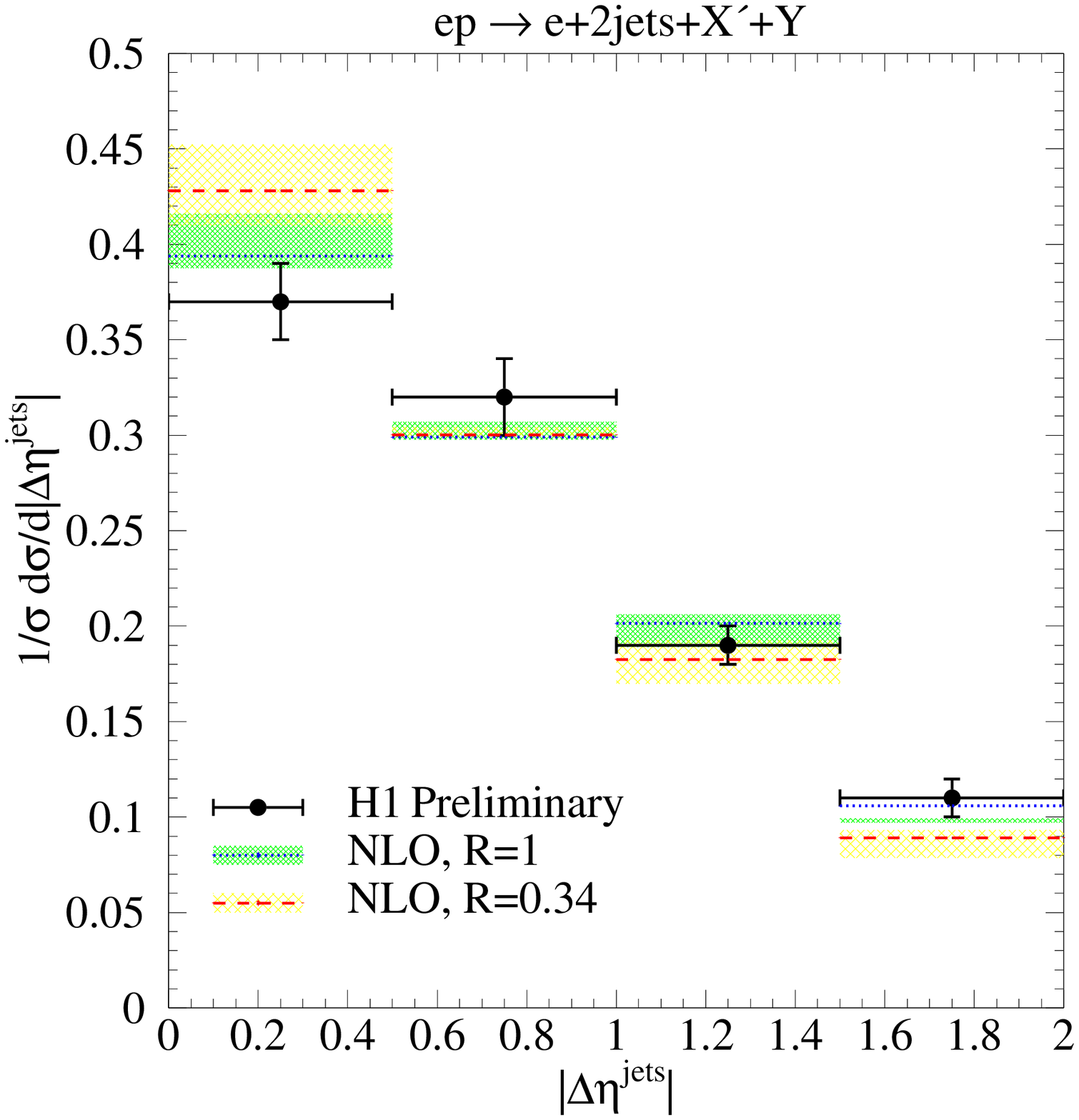,width=0.49\textwidth}
 \caption{\label{fig:7}Normalized $\overline{\eta}^{\rm jets}$ (left) and
 $|\Delta\eta^{\rm jets}|$ (right) distributions in LO (top) and NLO
 (bottom), compared to preliminary H1 data.}
\end{figure*}

Our LO (top) and NLO (bottom) results are shown in Fig.~\ref{fig:3} for
%
%
the differential cross sections in $x_{\gamma}^{\rm jets}$ (left) and
$z_{\p}^{\rm jets}$ (right), which are not normalized to one. The normalized
distributions in $x_{\gamma}^{\rm jets}$, $z_{\p}^{\rm jets}$, $\log_{10}
x_{\p}$, $y$, $E_T^{\rm jet1}$, $M_X^{\rm jets}$, $M_{12}^{\rm jets},
\overline{\eta}^{\rm jets}$, and $|\Delta\eta^{\rm jets}|$ are shown in LO
and NLO in Figs.\ \ref{fig:3a}-\ref{fig:7}.

For $d\sigma/dx_{\gamma}^{\rm jets}$ (Fig.~\ref{fig:3}, left), we have
very different cross sections for $R=1$ and $R=0.34$ and for the scale
choice $\xi = 1$. An
exception is the highest $x_{\gamma}^{\rm jets}$-bin, where the difference
is only $20\%$, since in this bin the direct contribution is dominant and
the suppression factor is therefore less effective. In all the other
bins, $d\sigma/dx_\gamma^{\rm jets}$ with $R=0.34$ is reduced by this factor
as expected. Except for the highest $x_{\gamma}^{\rm jets}$-bin, neither of
the two LO calculations agrees with the data. The $R=1$ cross section is too
large and the $R=0.34$ cross section is too small. Only when we consider the
scale variation with $0.5 \leq \xi \leq 2$ as a realistic error estimate,
we would conclude that the unsuppressed LO cross section ($R=1$) is
marginally consistent with the H1 data inside the experimental errors. At
NLO, the conclusion is reversed: the suppressed cross section now agrees
very well with the data, while the unsuppressed cross section drastically
overestimates the data.

For $d\sigma /dz_{\p}^{\rm jets}$ in Fig.~\ref{fig:3} (right), the agreement
of unsuppressed and suppressed cross sections with the data is equally
marginal at LO, even within the respective error bands, while it is
excellent for the suppressed NLO cross section. We remark that the
suppressed and unsuppressed cross sections with $\xi=1$ differ approximately
only by a factor 0.5, since in this distribution the direct and resolved
contributione are superimposed differently than in $d\sigma / dx_{\gamma}
^{\rm jets}$.

For the normalized $x_\gamma^{\rm jets}$ distributions in Fig.~\ref{fig:3a}
%
%
(left), the overall agreement is, of course, better. In particular, the
unsuppressed LO distribution agrees now with the data within the scale
uncertainty, whereas at NLO it is again the suppressed distribution that
describes the data best. Furthermore, the scale uncertainty is substantially
reduced in the normalized distributions as expected. For the $z_{\p}^{\rm
jets}$ distributions in Fig.~\ref{fig:3a} (right), both the unsuppressed
and suppressed LO distributions agree with the data within errors, while at
NLO agreement is only found for the latter.

The comparison of the normalized distributions in $\log_{10}$ $x_{\p}$ and
$y$ is shown in Fig.~\ref{fig:4}. Here the theoretical predictions for
%
%
$R=0.34$ and $R=1$ differ very little. This is understandable, since the
$x_{\p}$ and $y$ dependence of the cross section factorize (see Eq.\
(\ref{eq:13})) to a large extent. Only through the correlations due to the
kinematical constraints we observe small differences between the $R=0.34$
and the $R=1$ cross sections, particularly in the $y$ distribution. From
this comparison no definite conclusions concerning the suppression can be
drawn. All theoretical predictions agree more or less with the data. In the
highest $\log_{10} x_{\p}$ bin the measured point lies higher than the
theoretical points. This can be explained, at least partly, by an additional
sub-leading Reggeon contribution, which has not been taken into account in
the diffractive PDFs we are using (see Fig.~7 in \cite{H13}).

Next we look at the $E_T^{\rm jet1}$ distribution in Fig.~\ref{fig:5}. The
%
%
LO (left) and NLO (right) distributions with $R=0.34$ are flatter than the
unsuppressed distribution as we expect it, since the resolved component
occurs dominantly at the smaller $E_T^{\rm jet1}$. The suppressed cross
section agrees better with the data points, even if the scale uncertainty is
taken into account. Due to the normalization of the cross section, the
differences between LO and NLO are almost invisible.

The distributions $1/\sigma ~ d\sigma/dM_X^{\rm jets}$ and $1/\sigma ~
d\sigma/dM_{12}^{\rm jets}$ are correlated due to $M_X^{\rm jets}=M_{12}
^{\rm jets}/\sqrt{z_{\p}^{\rm jets}x_{\gamma}^{\rm jets}}$. Although the
distributions in $x_{\gamma}^{\rm jets}$ and $z_{\p}^{\rm jets}$ are bound
to reveal more detailed information on possible factorization breaking, we
have calculated the mass distributions nevertheless. The results and the
comparisons with the data are shown in Fig.~\ref{fig:6}. The
%
%
experimental cross sections increase with $M_X^{\rm jets}$, while they
decrease with increasing $M_{12}^{\rm jets}$. This is due to the correlation
mentioned above. The distribution in $M_{12}^{\rm jets}$ is also correlated
with the distribution in $E_T^{\rm jet1}$. For the mass distribution of the
dijet final state, which can directly be measured experimentally, the LO and
NLO, suppressed and unsuppressed distributions are very similar and agree
with the data. In contrast, the hadronic mass $M_X^{\rm jets}$ has
to be reconstructed and is very sensitive to systematic errors in the
measured variables. The theoretical prediction follows the increase in the
data only in LO, while at NLO the dependency is reversed and is very
sensitive to the presence of a possible third parton in the final state $X$.

The distributions in $\overline{\eta}^{\rm jets}$ and $|\Delta\eta^{\rm
jets}|$ presented in Fig.~\ref{fig:7} involve a delicate superposition of
%
%
direct and resolved contributions. In particular, the direct (resolved)
process dominates for negative (positive) $\overline{\eta}$. While the LO
$\overline{\eta}$-distri\-bution agrees better with the data, if the
resolved process is not suppressed ($R=1$), the conclusion is again reversed
at NLO, as was already the case for the $x_\gamma^{\rm jets}$-distribution
in Fig.~\ref{fig:3}. For the lowest bin in $\overline{\eta}$, we observe an
excess of the theoretical prediction over the data, which is well known from
studies of inclusive jet production at the very low transverse momenta
studied here and which can be related to additional hadronization effects.
The distribution in $|\Delta\eta^{\rm jets}|$ is intimately linked to the
angular distribution of the partonic scattering matrix elements. It is
thus less sensitive to the superposition of direct and resolved photon
contributions, and the theoretical predictions agree almost equally well.

In summary, we conclude that for most LO distributions the {\em
unsuppressed} theory, {\it i.e.} with no factorization breaking, agrees
better with the experimental data. This conclusion is, however, premature,
since at NLO it is the {\em suppressed} theory, {\it i.e.} with
factorization breaking and $R=0.34$, which is preferred.

In \cite{KKMR}, the suppression factor of $R=0.34$ was deduced from a
calculation of the ratio of diffractive and inclusive dijet photoproduction
at HERA as a function of $x_{\gamma}$ for two cases: (i) no absorption and
(ii) absorption included. The calculation of this ratio for the two cases
was based on a very simplified model, in which the ratio depended only on
the gluon PDFs of the pomeron and proton in the numerator and denominator,
respectively. It is of interest to see how this ratio behaves as a function
of $x_{\gamma}^{\rm jets}$ for the two cases $R=1$ and $R=0.34$ in LO and
NLO in the more detailed theory presented in this work, {\it i.e.} in a
theory where this ratio is calculated from the full cross section formula in
Eq.\ (\ref{eq:13}) and the corresponding formula for the inclusive dijet
cross section with quarks and gluons and realistic experimental cuts.

\begin{figure*}
 \centering
 \epsfig{file=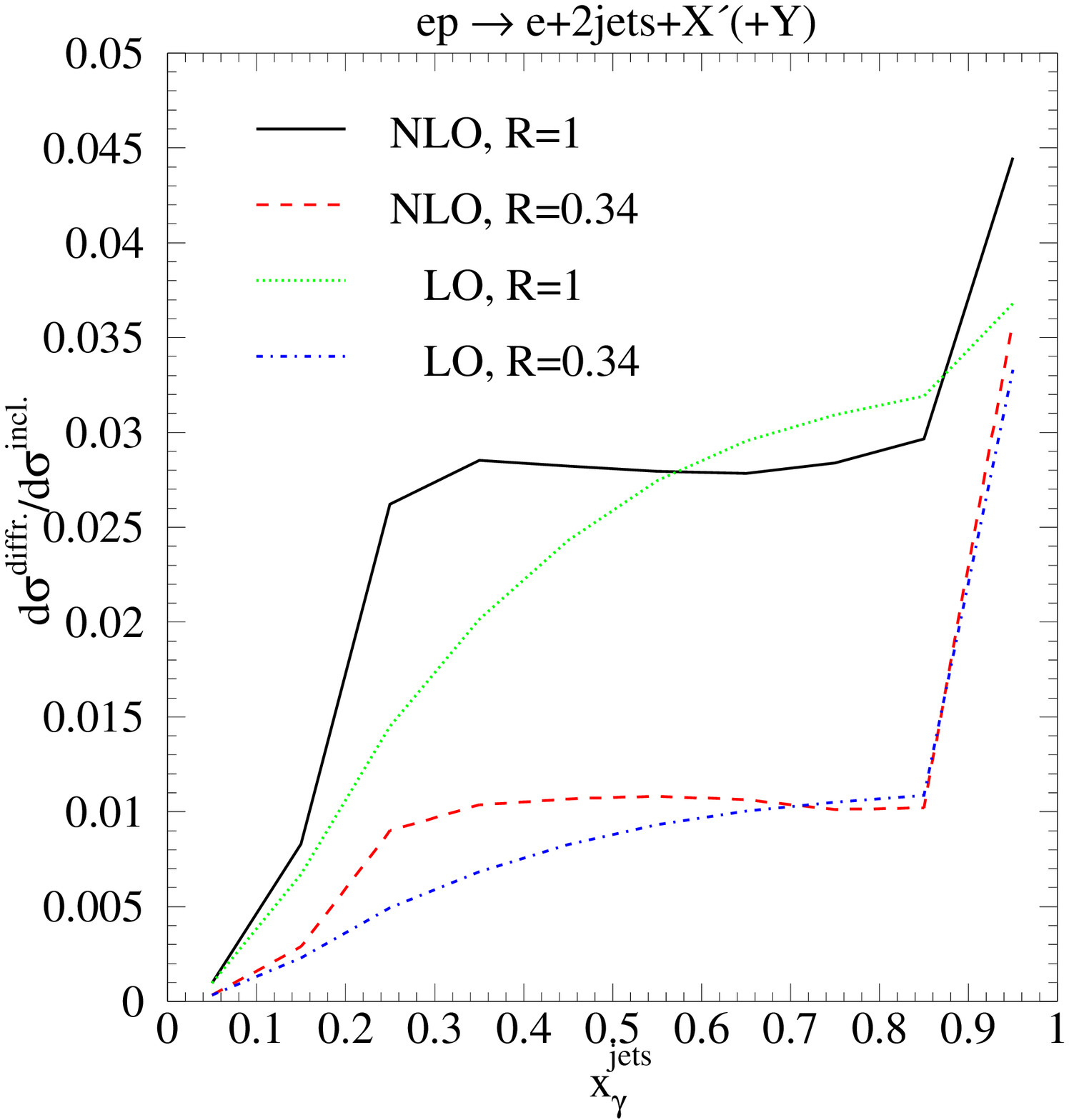,width=0.49\textwidth}
 \epsfig{file=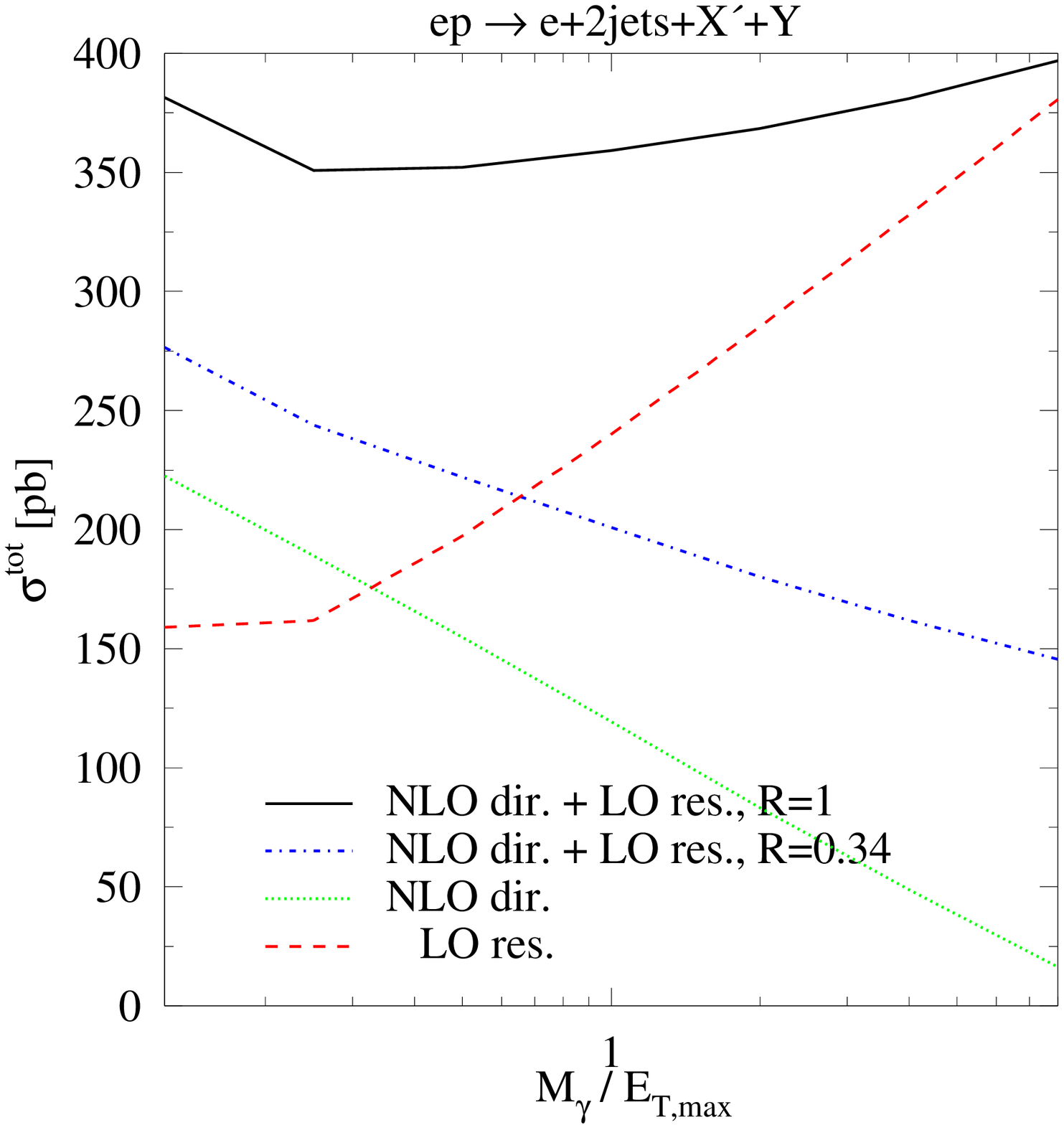,width=0.49\textwidth}
 \caption{\label{fig:8}Ratio of diffractive to inclusive dijet
 photoproduction as a function of $x_{\gamma}^{\rm jets}$ (left) and
 photon factorization scale dependence of resolved and direct contributions
 together with their weighted sums (right).}
\end{figure*}

The result is shown in Fig.~\ref{fig:8} (left), where we have used the
%
%
CTEQ5M1 parameterization for the proton PDFs \cite{CTEQ} in the inclusive
cross section results. In LO and for $R=1$, the ratio $d\sigma^{\rm diffr}/
d\sigma^{\rm incl}$ starts at small $x_{\gamma}^{\rm jets}=0.05$ at a very
low value ($\simeq 0.001$) and then rises monotonically up to $0.032$ and
$0.037$ at $x_{\gamma}^{\rm jets}=0.85$ and $0.95$. With $R=0.34$,
{\it i.e.} with suppression of the resolved part, the increase of this ratio
is very much reduced. It goes up to $0.011$ at $x_{\gamma}^{\rm jets}=0.85$.
At $x_{\gamma}^{\rm jets}=0.95$ the ratio is substantially larger, since in
this region the unsuppressed direct cross section dominates. We see that up
to $x_{\gamma}^{\rm jets}=0.85$ the suppressed ratio ($R=0.34$) is reduced
approximately by a factor of three as compared to the unsuppressed ratio
($R=1$) as expected. The behavior of the ratio is somewhat different for the
NLO case. In particular, the diffractive NLO resolved contribution has a
steeper rise at $x_{\gamma}^{\rm jets}=0.25$ and flatter behavior above,
which is reflected in both the unsuppressed and the suppressed sum. Compared
to the corresponding curves for $d\sigma^{\rm diffr}/d\sigma^{\rm incl}$
in \cite{KKMR}, the qualitative behavior of our curves, in LO and NLO, is
similar. The 'no absorption/absorption included' curves in \cite{KKMR}
resemble more our LO than our NLO results as expected. We have to keep in
mind, however, that the kinematic constraints applied in \cite{KKMR} differ
from ours, which are the same as in the experimental analysis. This
translates mainly into a different (smaller) normalization of our results.
Clearly it would be interesting to measure $d\sigma^{\rm diffr}/d\sigma
^{\rm incl}$ as a function of $x_{\gamma}^{\rm jets}$ in order to have
another observable for measuring the suppression as a function of
$x_{\gamma}^{\rm jets}$. Compared to the cross section $d\sigma/dx_{\gamma}
^{\rm jets}$ considered earlier, this ratio has the advantage to depend less
on the photon PDFs, which appear both in the numerator and the denominator
and should cancel to a large extent.

It may well be that our procedure to describe the factorization breaking by
applying a suppression factor to the total resolved cross section is not
correct and must be modified. An indication for this is the fact that the
separation between the direct and the resolved process is not physical. It
depends in NLO on the factorization scheme and scale $M_{\gamma}$, as
already mentioned earlier. The sum of both cross sections is the only
physically relevant cross section, which is approximately independent of the
factorization scale $M_{\gamma}$. By multiplying the resolved part with the
suppression factor $R=0.34$ the correlation of the $M_{\gamma}$-dependence
between the direct and the resolved part is changed and the sum of both
parts has a much stronger $M_{\gamma}$ dependence than for the unsuppressed
case ($R=1$). This is shown in Fig.~\ref{fig:8} (right). We see the
compensation of the $M_{\gamma}$-dependence between the NLO direct cross
section (dotted line) and the LO resolved cross section (dashed line) in the
unsuppressed ($R=1$) case, leading to a fairly $M_{\gamma}$ independent sum
of both contributions (full line) \cite{KKK,BKS}. When the LO resolved part
is suppressed with the factor $R=0.34$, the compensation is reduced, and the
sum of the NLO direct and LO resolved parts becomes much more
$M_{\gamma}$-dependent than before (although not too much in the range
$0.5 < M_{\gamma}/E_{T,\max} < 2$, as seen by the dashed-dotted curve in
the right part of Fig.~\ref{fig:8}).

The compensation of the $M_{\gamma}$-dependence between the NLO direct and
LO  resolved cross section occurs via the anomalous or point-like part of
the photon PDFs. This means that this part of the PDFs is closely related to
the direct cross section. It is usually assumed that the direct part obeys
factorization and has no suppression factor. So the point-like part in the
photon PDFs should not be suppressed either, and the suppression factor
should be applied only to the hadron-like part and the gluon part of the
photon PDFs. Since all three parts, point-like, hadron-like and gluon, are
correlated through the evolution equations, it is not clear how this
suggestion could be realized. Of course, if the point-like part is not
suppressed, the problem of the insufficient compensation of the scale
dependence of the NLO direct and LO resolved part would be solved. It is,
however, conceivable that this problem would be solved quite naturally if
one attempts to incorporate absorptive effects into the NLO theory
following, for example, the work of \cite{Kaidalov:2001iz}.

\section{Conclusions and Outlook}
\label{sec:4}

The recent measurement of diffractive dijet photoproduction combined with
the analysis of diffractive inclusive DIS data in terms of diffractive PDFs
offers the opportunity to test factorization in diffractive dijet
photoproduction. For this purpose we have calculated several cross sections
and normalized distributions for various kinematical variables in LO and NLO
and compared them with recent preliminary H1 measurements \cite{H13}. In LO
we found that the measured distributions und unnormalized cross sections
agree quite well with the theoretical results if, by a reasonable variation
of scales, a theoretical error is taken into account. This means that in a
LO comparison there is no evidence for a possible factorization breaking
expected for the resolved contribution. However, it is well known that for
dijet photoproduction NLO corrections are very important for the direct and
in particular for the resolved contributions to the cross section. Indeed,
the theoretical results at NLO disagree with the data for unnormalized cross
sections like $d\sigma/dx_{\gamma}^{\rm jets}$ and $d\sigma/dz_{\p}^{\rm
jets}$. Agreement between data and theoretical results is found, however, if
the resolved contribution is suppressed by a factor $R=0.34$. This factor is
motivated by a recent calculation of absorptive effects in diffractive dijet
photoproduction \cite{KKMR}. Since NLO results are more trustworthy than any
LO cross section calculations, we consider our findings a strong indication
that factorization breaking occurs in diffractive dijet photoproduction with
a rate of suppression expected from theoretical models.

It would be interesting to investigate hard diffractive photoproduction of
other final states, for which the superposition of direct and resolved
contributions is different. Such diffractive photoproduction reactions are,
{\it e.g.}, large-$p_T$ inclusive single-hadron production, heavy-flavour
production with or without jets, and prompt photon production. In order to
verify that factorization breaking disappears when the $Q^2$ of the virtual
photon is increased from small to larger values, it would be desirable to
have measurements of diffractive production of the final states mentioned
above as a function of $Q^2$.

Finally factorization breaking is expected not only in the diffractive
region, $x_{\p} \ll 1$, but also at larger values of $x_{\p}$ where Regge
exchanges other than the pomeron occur. For example, pion exchange is strong
in all reactions with a leading neutron. Here, dijet photoproduction with a
leading neutron has been studied in LO and NLO \cite{KKn} and compared to
ZEUS experimental data \cite{Zeusn}. This process could also be a candidate
for factorization breaking in the resolved contribution.

\begin{acknowledgement}
 This work has been supported by Deutsche Forschungsgemeinschaft through
 Grant No.\ KL 1266/1-3. We thank
 J.-M.\ Richard for a careful reading of the manuscript.
\end{acknowledgement}



\end{document}